\documentclass[reprint,twocolumn,floatfix,preprintnumbers,prb ,nofootinbib,hyperref]{revtex4-2}
\usepackage[T2A,T1]{fontenc}
\usepackage[utf8]{inputenc}
\usepackage[russian,english]{babel}
\usepackage{amsmath}
\usepackage{amssymb}
\usepackage{verbatim}
\usepackage{subcaption}
\usepackage{graphicx}
\usepackage{float}
\usepackage{booktabs}
\usepackage{wrapfig}
\usepackage[dvipsnames]{xcolor}
\usepackage{mathtools}

\usepackage[breaklinks=true,hidelinks]{hyperref}
\hypersetup{allcolors=[rgb]{0.0 0.0 0.6},linkcolor=[rgb]{0.75 0.05 0.05}}

\begin{document}  
\preprint{APS/123-QED}
\preprint{NORDITA-2022-013}
\title{Wire metamaterial filled metallic resonators}


\author{Rustam~Balafendiev}
\email{rustam.balafendiev@metalab.ifmo.ru}
\affiliation{School of Physics and Engineering, ITMO University, 197101 St. Petersburg, Russia}
\author{Constantin~Simovski}
 \email{konstantin.simovski@aalto.fi}
\affiliation{Department of Electronics and Nanoengineering,
Aalto University, School of Electrical Engineering, P.O. Box 13000, 0007
Aalto, Finland}
\author{Alexander~J.~Millar}
\email{alexander.millar@fysik.su.se}
\affiliation{The Oskar Klein Centre for Cosmoparticle Physics,
Department of Physics, Stockholm University,   Albanova University Center, SE-10691 Stockholm, Sweden}
\affiliation{Nordita, KTH Royal Institute of Technology and Stockholm University, Roslagstullsbacken 23, 10691 Stockholm, Sweden}

\author{Pavel~Belov}
\email{belov@metalab.ifmo.ru}
\affiliation{School of Physics and Engineering, ITMO University, 197101 St. Petersburg, Russia}
\date{\today}

\graphicspath{{./figures}}

\def \be {\begin{equation}}
\def \ee {\end{equation}}
\def \ve {\varepsilon}
\def\l#1{\label{eq:#1}}
\def\r#1{(\ref{eq:#1})}

\captionsetup{justification=raggedright,singlelinecheck=false}
\begin{abstract}
In this work we study electromagnetic properties of a resonator recently suggested 
for the search of axions --- a hypothetical candidate to explain dark matter. A wire medium loaded resonator (called a plasma haloscope when used to search for dark matter) consists of a
box filled with a dense array of parallel wires electrically connected to top and bottom walls. 
We show that the homogenization model of a wire medium works for this resonator
without mesoscopic corrections, and that the resonator quality factor $Q$ at the frequency 
of our interest drops versus the growth of the resonator volume $V$ until it is dominated by resistive losses in the wires. 
We find that even at room temperature metals like copper can give quality factors in the thousands --- an order of magnitude higher than originally assumed.
Our theoretical results for both loaded and unloaded resonator quality factors were confirmed by building an experimental prototype.  
We discuss ways to further improve wire medium loaded resonators.
\end{abstract}

\maketitle

\section{Introduction}

Despite wire media (WM) being one of the first metamaterials~\cite{Pendry:1998}, their full potential has yet to be explored. The unique properties of WM, an indefinite dielectric media with strong spatial dispersion, has led to it being important in novel applications from the radio frequencies to the optical range. 
Simple wire media (uniform arrays of parallel wires) offer strong anisotropy in one frequency range and nearly zero effective permittivity in another range, whereas 
double wire media grant amazingly high values of the permittivity. In the overview work~\cite{Sim} one may find a long list of WM applications known up to 2012 -- from far-zone transport of near field images to drastic enhancement of thermophotovoltaic generators. Since that time simple WM of semiconductor nanowires  have found applications in modern solar panels (so-called black silicon~\cite{Sim0}), microwave antennas with improved patterns~\cite{Forati2015}, and as frequency selective thermal emitters for radiative cooling of microlasers and in near-field thermophotovoltaic systems~\cite{Si,Li}. 

However, the use of WM in resonant systems has so far been very sparsely studied. One can mention only works~\cite{Sim1,Sim2}, where finite samples of nanowire media 
were suggested as radiating resonators granting a huge Purcell factor to fluorescence emitters in the mid-IR and near-IR bands, respectively.     
However, WM resonators are not only capable of enhancing molecular emission. One exciting possibility is to use a finite sample of WM in order to implement a novel type of microwave resonator called a {\it plasma haloscope} in the seminal work of Ref.~\cite{Lawson2019} to search for dark matter. The resonator suggested in this theoretical paper is a metal box filled with a simple wire medium. The governing idea of this resonator is based on the behavior of the WM as an effective plasma. 

At its plasma frequency the effective permittivity of WM crosses zero, which enables the epsilon-near-zero (ENZ) regime greatly increasing the wavelength of light in the medium. In contrast to the case of optical frequencies, where a number of natural materials such as semiconductors and metals can be used for ENZ, at radio frequencies this can only be achieved with artificial materials, and the WM was chosen in~\cite{Lawson2019} as a straightforward candidate. The behavior of WM in the ENZ regime has been partially studied: microwave WM in this regime offered  energy tunneling through subwavelength channels~\cite{Silveirinha2006} and 
excellent directionality of antennas~\cite{Alu2007, Zhou2010, Forati2015}, whereas infrared WM granted the unbounded spatial spectrum of eigenmodes 
promising the giant enhancement of Raman radiation and fluorescence of molecules located in such media~\cite{Sim3}. However, the idea of Ref.~\cite{Lawson2019} exploited the increase of the effective wavelength that enables at a given frequency a very low mode number for a very big resonator, much bigger than the empty metal cavities having the fundamental mode at the same frequency.  The low-frequency resonance for a big resonator together with the high uniformity of the eigenmode field (at much larger scales than a traditional cavity allows) makes the plasma haloscope promising for the search of axions, hypothetical particles of dark matter.  

The composition of dark matter is one of the most enduring problems in modern cosmology. While historically considered the second most promising candidate, the axion has risen in prominence due to the continued non-detection of weakly interacting massive particles. Originally proposed to solve the Strong CP problem, the mysteriously precise conservation of Charge-Parity symmetry in the Strong interaction~\cite{Peccei:1977hh,Weinberg:1977ma,Wilczek:1977pj}, the axion can be produced non-thermally in the early Universe to provide a natural dark matter candidate~\cite{Preskill:1982cy,Abbott:1982af,Dine:1982ah,Bergstrom:2000pn,Jaeckel:2010ni,Feng:2010gw}.

Due to an anomalous coupling to electromagnetism, the axion mixes with photons when a strong external magnetic field is applied. This mixing generates a small electric field, oscillating at a frequency corresponding to the axion mass and spatially constant on scales smaller than the de Broglie wavelength (which, due to the small escape velocity of the galaxy is approximately 1000 times larger than the Compton wavelength). Traditionally, this effect has been used to search for axions via the excitation of cavities~\cite{Sikivie:1983ip}.  
The coupling between the axion and the experiment is determined by the overlap of cavity mode with the axion field, with the highest overlap occurring for the most homogeneous mode~\cite{Sikivie:1983ip}. Since the exact value of the axion mass is unknown, the experimental setup has to be tunable over a wide range of frequencies, as well as being large enough to provide a reasonable signal, with power scaling with the volume and quality factor of the resonator~\cite{Sikivie:1983ip}.

While making a large microwave resonator is straightforward at the frequencies on the order of hundreds of MHz to GHz~\cite{Sikivie:1983ip,Rybka:2014xca,Woohyun:2016}, the axion may have significantly higher masses, with some recent calculations predicting a frequency of $16\pm 1.5$~GHz~\cite{Buschmann:2021sdq}. At these frequencies a conventional resonator has dimensions of a few centimeters, leading to both a massive loss of signal power and increased mechanical complexity. As a result, novel experimental methods are needed, with recent proposals using multiple or coupled cavities~\cite{Goryachev:2017wpw,Melcon:2018dba,Melcon:2020xvj,Jeong:2020cwz}, cavities with dielectric inserts to modify the mode structure~\cite{Carosi:2020akt,Quiskamp:2020yrx, Kim2020, Alesini2020, Alesini2021}, or even abandoning a traditional resonator in favor of a mirror~\cite{Horns:2012jf,Jaeckel:2013sqa,Suzuki:2015sza,Experiment:2017icw,BREAD:2021tpx} or of an array of large dielectric disks \cite{TheMADMAXWorkingGroup:2016hpc,Baryakhtar:2018doz, Chiles:2021gxk,   manenti2021search}. 

As shown in Ref.~\cite{Lawson2019}, the plasma haloscope allows one to overcome the mass difference between the axion and the photon, allowing for much larger resonant systems with close to homogeneous mode structures. While natural plasma operable at cryogenic temperatures with low loss and a controllable plasma frequency in the microwave regime does not exist, WM allow for a bespoke plasma to be made. WM made of a non-magnetic metal such as copper can operate in high DC magnetic fields and at low temperatures. They exhibit low loss, and, importantly, can be controlled mechanically (e.g., moving the wires) to allow simple tuning mechanisms.

However, previous work (Refs.~\cite{Lawson2019,Caputo:2020quz,Gelmini:2020kcu}) were based on a simple effective medium approach, and did not calculate the properties of such a medium from first princples. First of all, WM are spatially dispersive, and it is not evident 
how the effective permittivity known for an unbounded WM is applicable to a finite sample. Is it mesoscopic (sensitive to the sample sizes and to the surroundings, 
i.e., the walls being metal or not)? 
Mesocopy was not considered in Refs.~\cite{Lawson2019,Caputo:2020quz,Gelmini:2020kcu} but it is well known that mesoscopy exists for a layer of low-loss WM located in a dielectric host or in free space. In this situation the effective permittivity of WM is useless without the so-called additional boundary conditions (ABCs) \cite{Mario}. The problem of boundary conditions in metamaterials can be highly nontrivial as evidenced, for instance, by the recent rigorous analysis of boundary conditions in layered metamaterials~\cite{PhysRevB.101.075127}. 
While in Refs.~\cite{Sim1,Sim2} finite samples of WM were successfully modeled without ABCs, this was possible only because they were lossy (as was shown in work \cite{Sajjad}, for lossy WM layers ABCs are not needed). In our case, the WM sample must have low losses. Thus, the applicability of the homogenization model has to be checked in view of the possible mesoscopy.
Moreover, Refs.~\cite{Lawson2019,Caputo:2020quz,Gelmini:2020kcu} assumed that the volume of the system $V$ and quality factor $Q$ were unrelated, an important assumption as the signal strength of an experiment would be proportional to $QV$. Wires are not perfectly conducting and by filling the cavity with them we bring significant losses into the resonator. This effect can be taken into account by a homogenous effective medium approach, but the size of the effect has not been estimated. Also, there may be additional losses at the places where the wires are connected to the cavity walls. This possible effect is not taken into account by the homogenization model. Together, the WM sample inside the cavity must downgrade its quality factor, but by how much? How does this downgrading counteract the gain granted by the enlarged volume $V$?

In our work, we estimate these issues, providing a comparison of the analytical model with full-wave simulations and measurements. 
Our results are positive: we demonstrated the feasibility of a large-scale haloscope. We analytically derived the quality factor of the WM loaded cavity and 
made analytical calculations of the mode frequencies and mode fields. Comparison with the full-wave simulations has shown  
the analytic model works without any mesoscopic corrections if the metal walls are separated from the wires by one half of the WM internal period. We found that 
even at room temperatures one can have $Q$ by an order of magnitude higher than that heuristically assumed in Refs.~\cite{Lawson2019,Gelmini:2020kcu}.  Namely, using rather thick wires of polished copper we may reach quality factors higher than 3000, whereas further improvement may be granted by cryogenic environment. We have validated the analytic model not only by numerical simulations but also by building and measuring a $10\times 10$ wire array in a cubic metal box. Our numerical and experimental results are in good agreement and pave the way for further studies of large, controllable plasma haloscopes as advantageous setups for the search for dark matter.

\section{Analytical model of the resonator}
\label{analytic}
At the most basic level, a WM resonator consists of an array of wires, which in this work we will consider to be encased inside a metallic cavity. Using a cylindrical cavity with a square lattice of wires results in the variable distance between the edge of the metamaterial and the wall. Depending on the wall spacing and how the cavity is filled, there is a possibility of eigenmodes being formed in the air gaps at frequencies comparable or lower than the plasma frequency. We consider a cubic cavity as it matches the lattice geometry, leading to a constant and simple distance between the cavity walls and wires. The walls ensure that the only losses in the system are either resistive losses or coupling to an external antenna, leading to a stronger resonance. To understand the general behavior of such a system, we will start with an analytic approach.

\begin{figure}[h]
\centering
    \begin{subfigure}[c]{.44\linewidth}
\includegraphics[width=\linewidth]{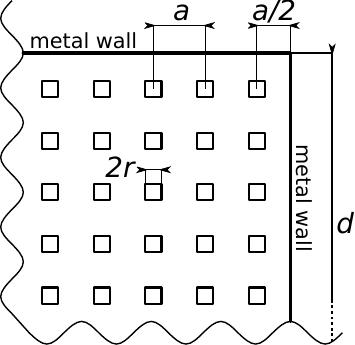}
\caption{}\label{fig:wm_res_2d}
\end{subfigure}
\begin{subfigure}[c]{.54\linewidth}
\includegraphics[width=\linewidth]{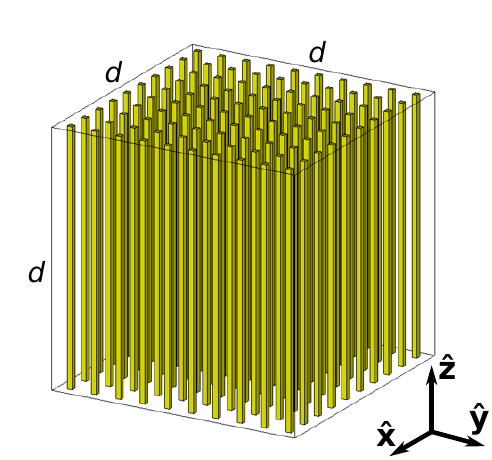}
\caption{}\label{fig:wm_res_3dsub}
\end{subfigure}
    \caption{{\em Left:} The corner of the $xy$ cross-section for a resonator consisting of an array of square wires of thickness $2r$, with spacing $a$ with a gap of $a/2$ between the wire array and an enclosing metal cavity of side length $d$. {\em Right:}  3D model of the $10 \times 10$ example  wire medium cubic resonator which will be used in numerical simulations. The wires have a square cross-section for ease of calculation with a hexahedral mesh in CST MWS \cite{cst} Eigenmode solver. The number of wires allows for manageable computational requirements while operating in a regime where one expects the homogeneous effective medium description to be valid.}
    \label{fig:wm_res_3d}
\end{figure}

For simplicity we will take the WM to be a square grid of $N \times N$ metal wires of thickness $2r$ arranged with a period $a$ (as depicted in Fig.~\ref{fig:wm_res_3d}). We will consider wires with both rectangular and round cross-sections. Lateral walls of the cubic metal cavity are distanced by $a/2$ from the centers of the edge wires so that the side length $d$ of the resonator is equal to $d=N a$. The distance of $a/2$ was chosen since it is the distance by which the effective medium extends out of the material~\cite{Simovski2018}.
The wire length is defined as $d$ and the wires are terminated at the top and bottom walls.

 The dielectric permittivity tensor $ \hat{\varepsilon}$ of a simple WM composed of $z$-oriented wires is given by (see, e.g., \cite{Sim})
 \be{}
 \hat{\varepsilon} = \left(\begin{matrix}
 1 & 0 & 0\\
 0 & 1 & 0\\
 0 & 0 & \varepsilon_{zz}
 \end{matrix}\right)\,,
 \l{va0}
 \ee{}
where in the absence of losses we have:
\be{}
\varepsilon_{zz} = 1 - \frac{k_p^2}{k_0^2-k_z^2}\,,
\l{va}
\ee{}
where $k_0$ is the free space momentum ($\omega/c$) and $k_z$ is the momentum in the $z$ direction.
This gives a maximally anisotropic plasma, so that modes $E$-fields perpendicular to the wires "perceive" an almost empty cavity.
For the special case when the $E$-field is aligned strictly in the $z$-direction (i.e., along the wires) this formula becomes identical to that describing the permittivity of a 
non-magnetic collision-free plasma: 
\be{}
\varepsilon\equiv \left.\varepsilon_{zz}\right|_{k_z = 0} = 1 - \left(\frac{\omega_p}{\omega}\right)^2.
\l{va1}
\ee{}
In writing these we have written the effective plasma frequency as $\omega_p$ with $k_p = \omega_p/c$ the plasma wave number defined in \cite{Belov2002} for round wires as 
\be{}
k_p^2 = \frac{2\pi/a^2}{\ln{\frac{a}{2\pi r}} + F(1)}\,. \label{eq:wavenumber}
\ee{}
Here F is a function of the ratio of the periods in the lattice of wires for a rectangular array. For square lattices $F(1)=0.5275$~\cite{Belov2002}. As we only deal with the case of square lattices, we will write $F(1)\equiv F$.
In other words, the effective plasma frequency and wavenumber are simply given by the geometry of the system, depending only on $a$ and $r$. This allows for custom plasmas to be created with the plasma frequency desired for any particular application, or even tuned mechanically~\cite{Lawson2019,Gelmini:2020kcu}.  

\subsubsection*{Modes}
The eigenmodes of the whole structure (WM in the cavity) can be split into Transverse Electric (TE) modes with the magnetic field aligned with the wires and Transverse Magnetic (TM) modes with the electric field aligned instead.
For the TE modes, the $E$-field vectors lack any component along $z$ and if the wires are thin enough their quasi-static interaction with the wires is negligible. 
As a result, the frequencies of TE modes are nearly equal to the frequencies of an empty cavity. Because of this, we will focus primarily on the TM modes, however if the WM is extended to have wires in the $x,y$-directions we would expect a similar story to play out for the TE modes.

For the TM modes, the $E$-field is confined to the $z$ axis and so sees a plasma medium. Below the plasma frequency the effective refractive index is predominantly imaginary and so any  waves are evanescent. Near the plasma frequency, the WM enters an ENZ regime leading to significantly larger wavelengths. Because of this the TM modes are shifted to just above the plasma frequency, leading to the fundamental mode being much higher in frequency than would be realized for an empty cavity. 

Neglecting losses, by requiring that the $E$-fields satisfy the conducting boundary conditions at the walls, the TM modes $E_z^{lmn}$ with eigenfrequencies $\omega_{\rm res}^{lmn}$ of amplitude $E_0$ can be written as 
\begin{equation}
	E_z^{lmn}=E_0\cos(k_xx)\cos(k_yy)\cos(k_zz)\,,\label{eq:modes}
\end{equation}
with eigenfrequencies given by 
\begin{equation}
    \epsilon \left(\frac{\omega_{\rm res}^{lmn}}{c}\right)^2=\left (\frac{l\pi}{d}\right)^2+\left (\frac{m\pi}{d}\right)^2+\epsilon\left (\frac{n\pi}{d}\right)^2\,,
\end{equation}
where we have used that $k_x=l\pi/d,k_y=m\pi/d$, and $k_z=n\pi/d$. As one of the primary applications of resonant WM are plasma haloscopes, we will focus on the fundamental (TM110) mode. As this mode has the highest spatial homogeneity, it would couple most strongly to dark matter and provide the strongest signal~\cite{Lawson2019}. In this case, we will define $\omega_{\rm res}\equiv\omega_{\rm res}^{110}$, which allows us to write
\begin{equation}
\left(\frac{\omega_{\rm res}}{c}\right)^2 \equiv k_{\rm res}^2 =  k_p^2 + 2\left(\frac{\pi}{d}\right)^2.
\end{equation}

\begin{figure}[h]
\centering
    \includegraphics[width=\linewidth]{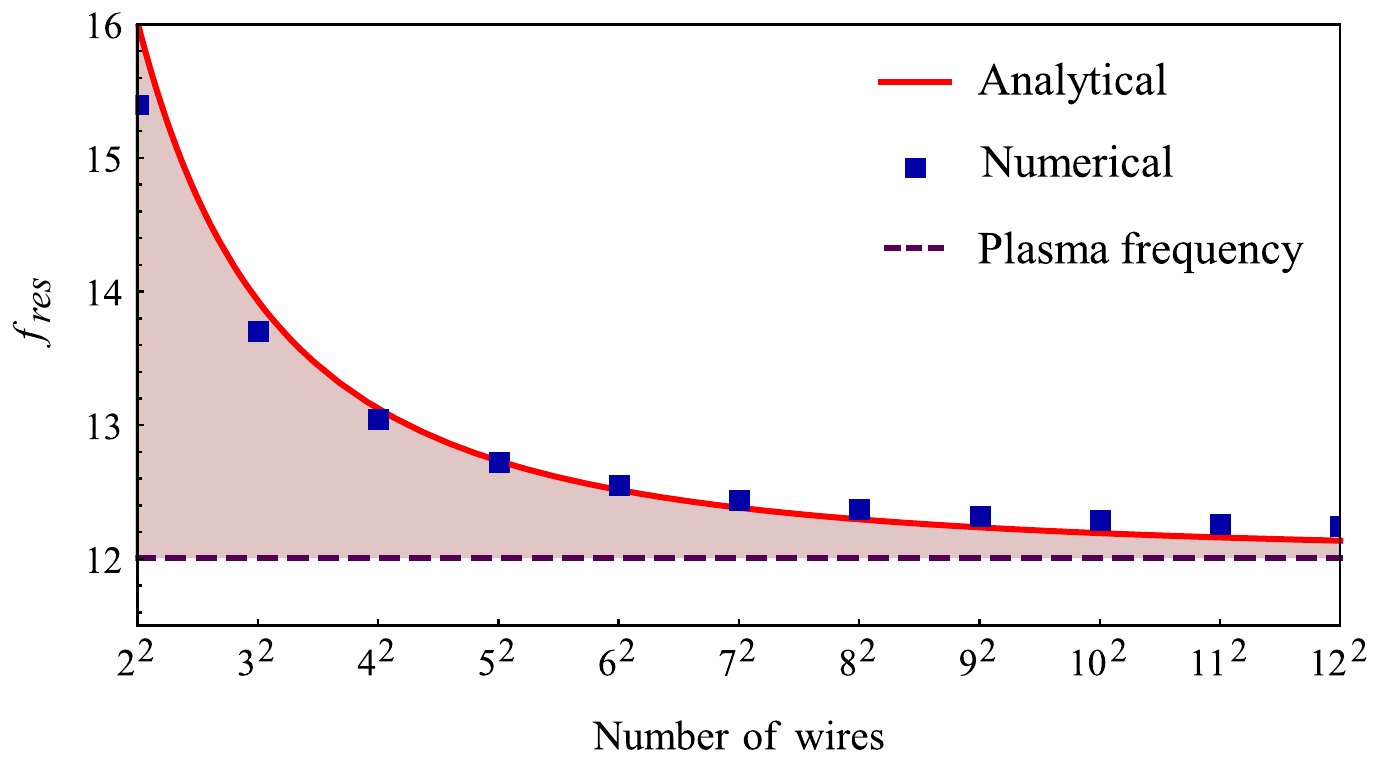}
    \caption{Resonance frequency $f_{\rm res}$ as a function of the number of wires $N^2$ in a WM consisting of radius $r=1$\,mm wires with a period $a=1$\,cm placed inside a cubic cavity. Analytic results (red line) assuming a homogenization model are compared with a numerical simulation (blue squares) in CST \cite{cst} with square wires of thickness $2r$, with a gap of $a/2$ with the cavity walls. For comparison, the plasma frequency of the WM is shown via the black dashed line. As $N$ increases $f_{\rm res}$ approaches the plasma frequency of the WM.}
    \label{fig:wm_res_freq}
\end{figure}

We thus have a simple analytic expression for the resonant frequencies of the resonant WM. To check this analytic effective medium treatment, we compare the analytic calculations to full-wave simulations in CST Microwave Studio \cite{cst} using the eigenmode solver in Fig.~\ref{fig:wm_res_freq}. By showing the resonant frequency versus the number of wires across the resonator ($N=d/a$) we see that the analytic formulas provide an excellent prediction for the full numerical treatment. The agreement is particularly good for larger systems (1\%), meaning that the analytical treatment should be appropriate for the large sizes considered for plasma haloscopes. We note, however, that this numerical model used a square cross-section of wires for computational simplicity and faster meshing. As will be shown later, modeling a cavity with round wires results in a small but noticeable difference in the quality factor and plasma frequency. We see also that as the size of the WM increases the $\omega_{\rm res}\to \omega_p$ asymptotically. With this understanding of the mode structure, we can turn our attention to the losses.

\subsubsection*{Losses}

Unfortunately, normal conductors will always have resistive losses, leading us to one of the most salient properties of any resonator: the quality factor $Q$. To estimate the losses of the system, we must augment the dielectric constant to include an imaginary component. For simplicity, we will consider the same fundamental TM110 mode; however, other modes of the system can be analyzed similarly. In this case, we can treat the medium as isotropic and use the results of Ref.~\cite{Maslovski2009} to write  
\begin{equation}
\varepsilon = \varepsilon' - j\varepsilon'' 
= 1 - \frac{k_p^2}{k^2-j \xi k}\,,
\end{equation}
with real and imaginary parts given by
\begin{equation}
\begin{aligned}
\varepsilon' &= 1 - \frac{k_p^2}{k^2-\xi^2}\,,\\
\varepsilon'' &= \frac{\xi k_p^2/ k}{k^2-\xi^2}\,.
\end{aligned}
\l{va2}
\end{equation}
The loss coefficient $\xi$ is derived in Ref.~\cite{Maslovski2009} 
\be{}
\xi = \frac{Z_w}{L_w}\sqrt{\varepsilon_0\mu_0}\,,\\
\l{xi}
\ee{}
with $Z_w$ and $L_w$ being the wire impedance and inductance per unit length, respectively. For wires with circular cross-sections, these quantities can be calculated from the magnetic permeability $\mu$ and conductance $\sigma$ of the wire material as \cite{Maslovski2009,Olyslager2005} 

\begin{subequations}
\begin{align}
L_w &= \frac{\mu_0}{2\pi}({\ln{\frac{a}{2\pi r}} + F}) \,,\\
Z_w &= \text{Re}\left(\frac{\sqrt{-j\omega\mu}}{\sqrt{\sigma}2\pi r} \frac{J_0 (\sqrt{-j\omega\mu \sigma}r)}{J_1 (\sqrt{-j\omega\mu \sigma}r)}\right)\,.\label{LwZw}
\end{align}
    \end{subequations}

The electromagnetic field of the fundamental mode of a cubic resonator ($d\times d\times d$), loaded by a dielectric whose complex permittivity has real part $\varepsilon'$ much larger than $\varepsilon''$, can be obtained from Eq.~\eqref{eq:modes} explicitly as
\begin{equation}
\left. 
\begin{aligned}
    H_x&=jE_0\sqrt{\frac{\ve_0\ve'}{2\mu_0}} \cos{\frac{\pi x}{d}}\sin{\frac{\pi y}{d}}\\
    H_y&=-jE_0\sqrt{\frac{\ve_0\ve'}{2\mu_0}} \sin{\frac{\pi x}{d}}\cos{\frac{\pi y}{d}} \\
    E_z&=E_0 \cos{\frac{\pi x}{d}}\cos{\frac{\pi y}{d}}
\end{aligned}
\right\}
-d/2 \leq x,y \leq d/2\,,
\label{eq:fields}
\end{equation}
where we have assumed that the medium is non-magnetic ($\mu=\mu_0$).
 The quality factor $Q$ of the resonator at the mode eigenfrequency can then be found via the stored energy $W_\text{stored}$ and the power of the losses in the walls, $P_\text{walls}$, and in the wires, $P_\text{wires}$,  as follows:
\begin{equation}
    Q = \frac{W_\text{stored}}{W_\text{lost}} = \frac{\omega W_\text{stored}}{P_\text{walls}+P_\text{wires}}\,.
\end{equation}
The equation for the stored energy density in a dispersive dielectric includes a correcting coefficient $\frac{\partial(\omega \varepsilon'(\omega))}{\partial \omega}$ and is thus written as~\cite{landau1995electrodynamics}
\begin{equation}
\begin{aligned}
    w_\text{stored} &= \frac{\mu_0 |H_t|^2}{4} + \frac{\varepsilon_0 S |E_z|^2}{4}\,,\\
    S &= \frac{\partial(\omega \varepsilon'(\omega))}{\partial \omega} = 1 + \left(\frac{\omega_p}{\omega_{\rm res}}\right)^2\,.
\end{aligned}
\end{equation}
In the limit of $\omega_{\rm res} \xrightarrow{} \omega_p$ the coefficient $S = 2$ and the $H$-fields vanish. Integrating $w_{\rm stored}$ over the resonator volume gives:
\begin{equation}
    W_\text{stored} = \iiint_{-d/2}^{d/2}w_{\rm stored} dV = \frac{\ve_0 d^3 E_0^2}{16 } (\ve{}'+S)\,.
\label{eq:wstored}
\end{equation}

Losses in the walls, $P_{\rm walls}$, can be found by integrating the tangential components of the magnetic field over the walls. Ignoring an overall phase and using that the currents are given by the cross product of the normal vector $\bf \hat n$ via ${\mathbf J}=\mathbf{\hat n}\times\mathbf{H}$ we get
\begin{subequations}
\begin{align}
    \mathbf{J}\big|_{|x|=\pm d/2} &= 
    E_0\sqrt{\frac{\ve_0\ve'}{2\mu_0}} \cos{\frac{\pi y}{d}}\mathbf{\hat z}\,,\\
    \mathbf{J}\big|_{|y|=\pm d/2} &=   E_0\sqrt{\frac{\ve_0\ve'}{2\mu_0}} \cos{\frac{\pi x}{d}}\mathbf{\hat z}\,,\\
    \mathbf{J}\big|_{|z|=\pm d/2} &=\pm E_0\sqrt{\frac{\ve_0\ve'}{2\mu_0}} \sin{\frac{\pi x}{d}}\cos{\frac{\pi y}{d}}\mathbf{ \hat x}\nonumber \\&~\, \, \, \, \, \pm  E_0\sqrt{\frac{\ve_0\ve'}{2\mu_0}}\cos{\frac{\pi x}{d}}\sin{\frac{\pi y}{d}}\mathbf{ \hat y}\,,
\end{align}
\end{subequations}
we can then integrate to find \cite{pozar2011microwave}
\begin{equation}
\begin{aligned}
P_{\rm walls} &= \frac{R_s}{2}\iint_{-d/2}^{d/2}|\mathbf{J}|^2dS \\
&= \frac{3 R_s\ve_0\ve'}{4\mu_0}d^2 E_0^2,
\end{aligned}
\end{equation}
where $R_s = \sqrt{\omega\mu_0/2\sigma}$ is the surface resistivity of the cavity walls.

\begin{figure}[h]
\centering
    \includegraphics[width=\linewidth]{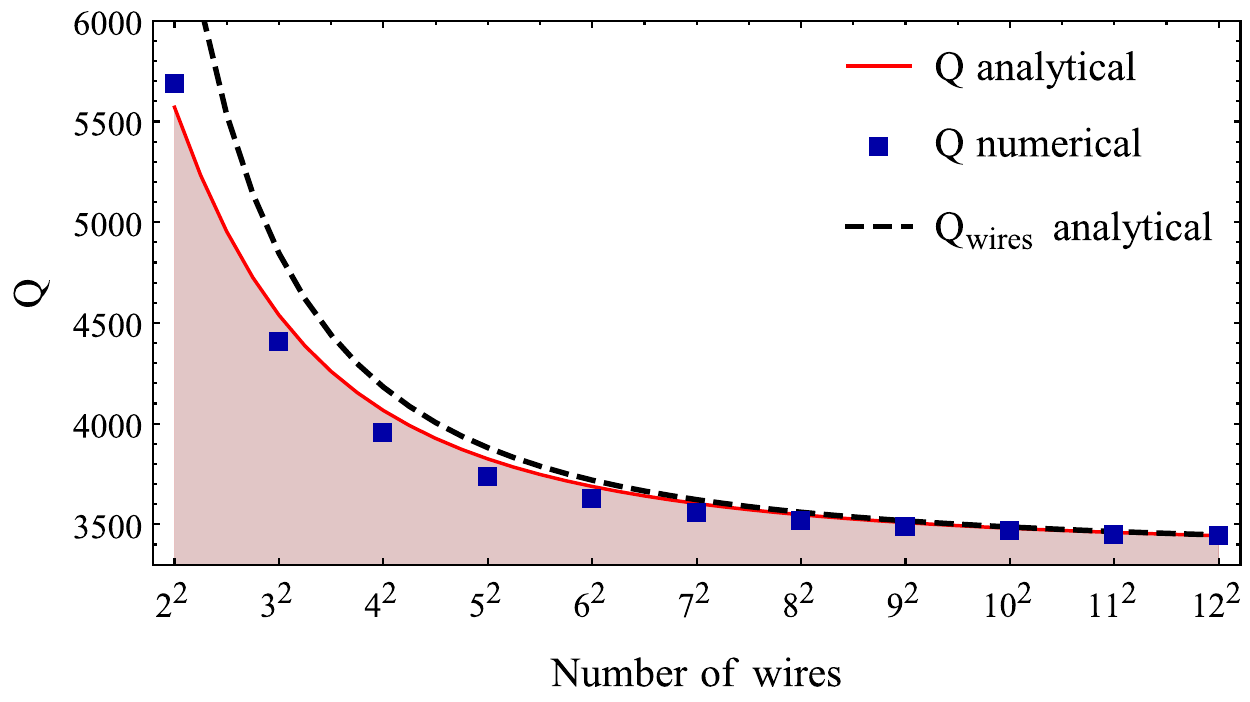}
    \caption{Unloaded quality factor $Q$ as a function of the number of wires $N^2$ in a WM consisting of radius $r=1$\,mm wires with a period $a=1$\,cm placed inside a cubic cavity. Analytic results (red line) assuming a homogenization model are compared with a numerical simulation (blue squares) in CST with square wires of thickness $2r$, with a gap of $a/2$ with the cavity walls. For comparison, the quality factor of the resonator with lossless walls $Q_{\rm wires}$ is shown (black dashed line). As $N$ increases, $Q$ asymptotically approaches that of an infinite wire medium (i.e., no wall losses) and $Q_{\rm wires}$ and $Q$ become virtually indistinguishable. In the limit of $N \to 0$, $Q_{\rm wires}$ approaches infinity, as there are no wires to cause resistive losses.}
    \label{fig:wm_res_Q}
\end{figure}

Similarly, the losses in the wires can be found by considering the imaginary component of the dielectric constant $\ve''$ using~\cite{landau1995electrodynamics}
\begin{multline}
P_{\rm wires}=\frac{\omega \ve{}_0 \ve{}''}{2}\iiint_{-d/2}^{d/2}|E_z|^2 dV = \omega\frac{\ve{}_0  \ve{}'' d^3 E_0^2}{8 } \,.
\label{eq:pwires}
\end{multline}
 As opposed to the losses in the walls, which grow with the surface area of the resonator, the losses in the wires grow proportionally to the cavity volume. As a result, as $d$ is increased, $P_{\rm wires}$ dominate over $P_{\rm walls}$ if the cavity is made of similar quality material. As axion experiments will focus on large volumes (and so large $d$) the wall losses can be mostly neglected. The resulting quality factor of the resonator can then be written using Eqs.~ (\ref{eq:wstored}) and (\ref{eq:pwires}) as:
\begin{equation}
    Q_{\rm wires} = \frac{\omega W_{\rm stored}}{P_{\rm wires}} = \frac{(\ve{}'+S)}{2\ve{}''} 
    \xrightarrow{k_{\rm res} = k_p} \frac{k_p}{\xi}-\frac{3}{2}\frac{\xi}{k_p}\,.
    \l{qq}
\end{equation}

As can be seen from Fig. \ref{fig:wm_res_Q} for a large resonant cavity in the case of a highly conducting metal ($\xi \ll k_p$), the quality factor 
depends on the size of the resonator in a manner similar to that of the mode frequency. Our analytic calculation is in excellent agreement with numerical simulations for all system sizes.
The worst agreement ($3^2$ wires with 2.4\%) provides a good match between analytics and simulations, improving to less than 1\% for $12^2$ and up. However, we can see that as the system size increases, the relative importance of wall losses decreases, with $Q\simeq Q_{\rm wires}$ for arrays larger than $\sim 7\times 7$, approaching an asymptotic value of $Q\simeq 3300$. Beyond this point, increasing the system size simply gives an almost linear increase in signal power, which is proportional to $QV$.
These analytical calculations were confirmed by full-wave simulations of square cross-section wires in CST MWS \cite{cst} Eigenmode Solver, whose results are shown in Fig.\ref{fig:wm_res_Q} as well. As the number of wires approaches the limit of a single wire $N=1$ both resonance frequency and Q-factor rise sharply, approaching the corresponding values calculated for an empty cubic metallic cavity with side length $d = a$. This can be explained by the losses in the walls starting to dominate over the losses in the medium in smaller WM-filled cavities. 

\begin{figure}[tb]
\includegraphics[width=1\linewidth]{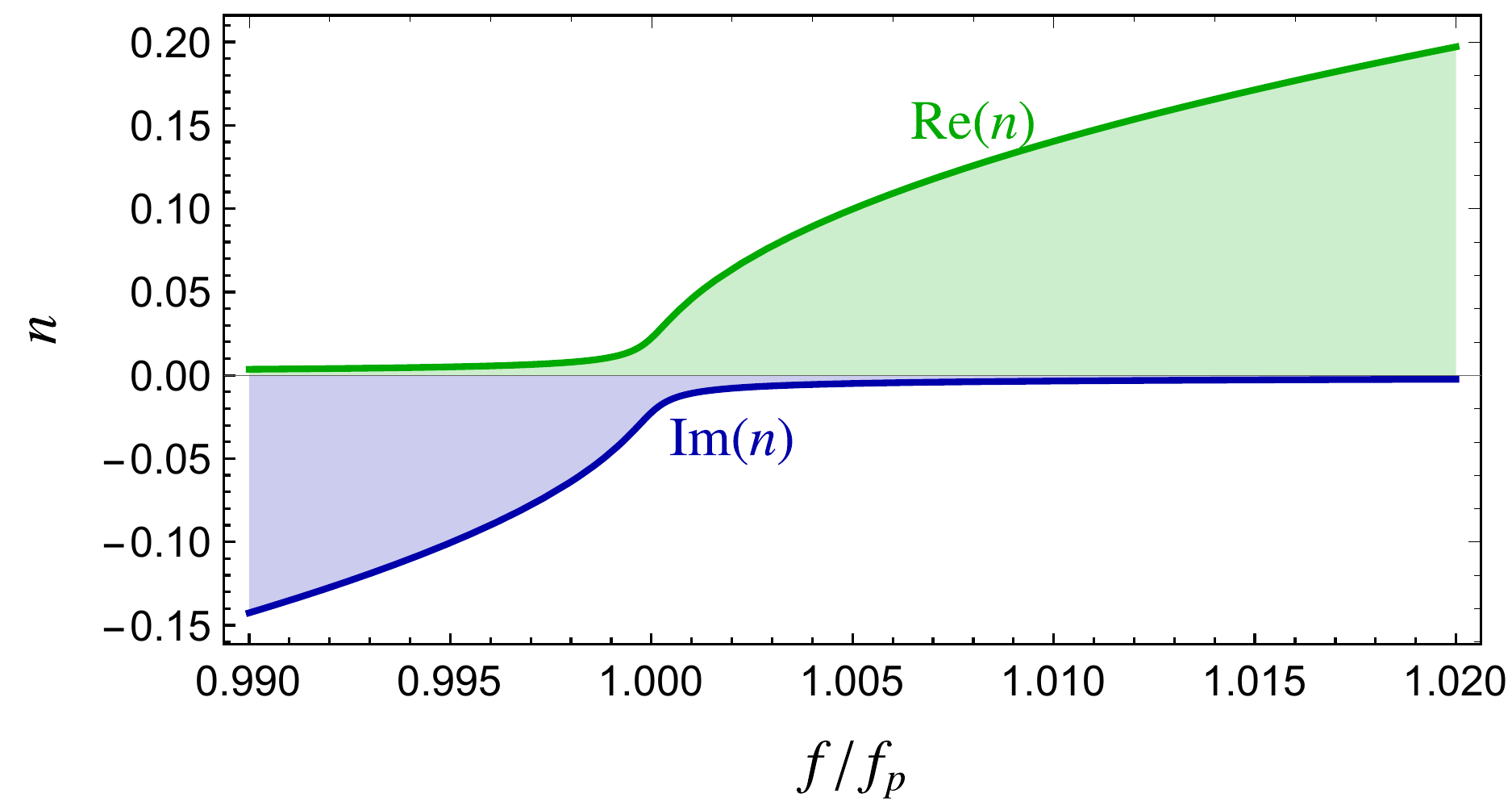}
    \caption{Real (green) and imaginary (blue) parts of the refractive index $n$ as a function of frequency $f$ around the plasma frequency $f_p$. For illustrative purposes, we have chosen a plasma with  $\xi=10^{-3}k_p$, i.e., a lifetime of 1000 cycles.}
    \label{fig:ref}
\end{figure}

While the quality factor remains constant for large system sizes, the existence of a single cavity mode actually breaks down. To see this, consider the refractive index $n=\sqrt{\ve\mu}$ near resonance, as shown in Fig.~\ref{fig:ref}. Near the plasma frequency the real and imaginary part of $n$ are equal up to a sign which depends on the time convention used (i.e., $e^{\pm i\omega t}$) with $\ve\simeq j/Q_{\rm wires} $ leading waves to decay with a characteristic length $\sim \sqrt {Q_{\rm wires}}\lambda_c$. This affects both the energy transport in the system, and the ability to form standing wave cavity modes. For cavities larger than this typical scale, waves will decay before they travel from one wall to another. In this case, our mode analysis breaks down. However, the analysis of Ref.~\cite{Lawson2019} does not assume that cavity modes form, and is thus unaffected. The axion would create an effective volume current, exciting the entire system uniformly. However, as the system would then be larger than the decay length,  multiple antennas would be required to pick up the full signal.   While we leave such considerations for future work, we note that the walls can be neglected and the quality factor of the cavity  will only depend on the power extracted by the antenna and the resistive losses of the wires, which will still be given by $Q_{\rm wires}$. 

Many applications of resonators rely on the resonance being high quality. With this in mind, how can one maximize the quality factor? In general, the losses will depend primarily on the thickness of the wires, as well as their conductivity. While one might anticipate that thicker wires would result in lower resistive losses in analogy to energy transport in power lines, this does not hold without limit. 
As the radius of the wires increases it starts to become comparable to the Compton wavelength, violating the assumption that the wire radius is much smaller than the wavelength (i.e., that the wires are thin).

 As shown above, for large resonators the wall losses can be neglected, so for simplicity we will treat the system as infinitely large (i.e., $f_{\rm res}=f_p$ and $Q=Q_{\rm wires}$). This allows us to choose an optimal wire radius for a given plasma frequency. To see how $Q$ scales with experimental parameters, we can explicitly write it as
\begin{equation}
    Q\simeq\frac{k_{p}}{\xi}=\frac{\sqrt{\omega_p\sigma\mu^{-1}}\mu_0 r}{{\rm Re}\left[\sqrt{-j}\frac{J_0 (\sqrt{-j\omega_p\mu \sigma}r)}{J_1 (\sqrt{-j\omega_p\mu \sigma}r)}\right]}\left(\ln{\frac{a}{2\pi r}}+F\right)\,.\label{eq:Q1}
\end{equation}
As we expect that larger wires will result in lower losses, we will focus on the regime where $r\gg 1/\sqrt{\omega\sigma\mu}$. In this limit the wires are much larger than the skin depth of the material $\delta=\sqrt{2/\omega\sigma\mu}$ and we can simplify Eq.~\eqref{eq:Q1} by noting that 
\begin{equation}
\frac{J_0 (\sqrt{-j}\alpha)}{J_1 (\sqrt{-j}\alpha)}\lim_{\alpha\to\infty}=j\,,
\end{equation}
giving
\begin{equation}
    Q\simeq\mu_0 r\sqrt{\frac{2\omega_p\sigma}{\mu}}\left(\ln{\frac{a}{2\pi r}}+F\right)=2\frac{\mu_0}{\mu}\frac{r}{\delta}\left(\ln{\frac{a}{2\pi r}}+F\right)\,.\label{eq:Q2}
\end{equation}
We see that $Q$ is determined by two factors. The first being given by the ratio of radius $r$ with the skin depth $\delta$ (up to the relative permeability, which for copper is close to unity). The second is a geometric term coming from the inductance $L_w$, which regulates $Q$ for high frequencies.  Note that $a$ is uniquely determined for a given $r$ and $\omega_p$, falling with increasing $\omega_p$ and increasing with larger $r$. We can rearrange Eq.~\eqref{eq:wavenumber} for the explicit form, given by
\begin{equation}
a=\frac{2\sqrt{\pi}}{\omega_p } {W}_0\left(\frac{e^{2F}}{\pi r^2\omega_p^2}\right)^{-1}\,,    
\end{equation}
where ${W}_0(x)$ is the principle branch Lambert $W$ function.

\begin{figure}[tb]
\includegraphics[width=1\linewidth]{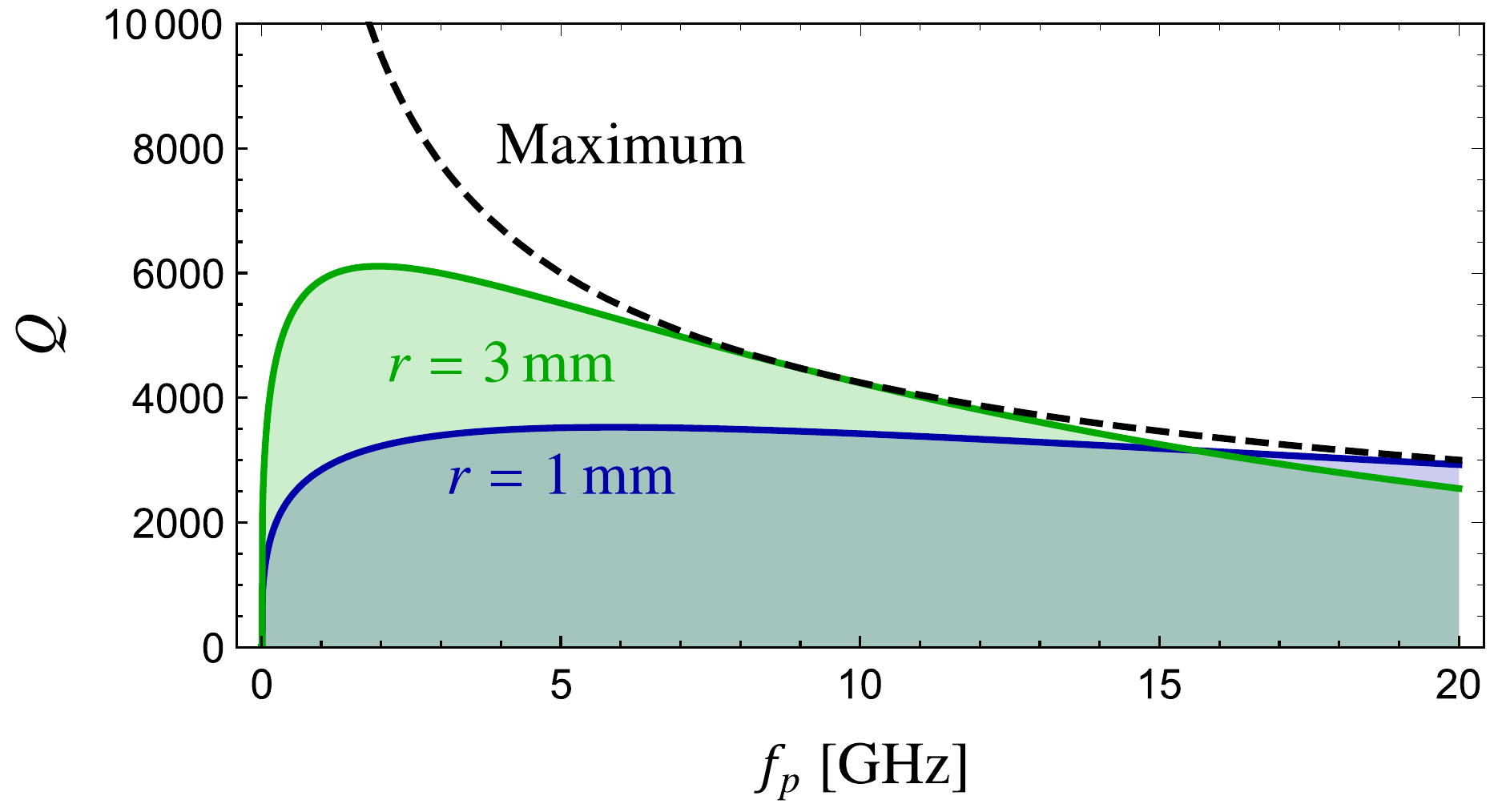}
    \caption{Unloaded quality factor $Q$ as a function of the plasma frequency $f_p$ for an infinite copper wire metamaterial at room temperature. We show wires with radius $r=1$~mm and $r=3$~mm in blue and green, respectively, with the black dashed line showing the maximum possible $Q$ for a given $f_p$. The period of the wires $a$ is adjusted so that the plasma frequency of the system is given by $f_p$. The theoretically maximum $Q$ diverges as one moves to DC as the optimal wire size and spacing approaches infinity.}
    \label{fig:radius}
\end{figure}

For some specific examples, we show $r=1,3\,$mm examples with $Q$ a function of $f_p=\omega_p/2\pi$ for room temperature copper in Fig.~\ref{fig:radius}. Due to the trade-off between decreasing the resistive losses and maintaining a thin wire limit, each wire radius has a maximum quality factor at a frequency which corresponds to $r\simeq\lambda_c/50$. However, the peak of the quality factor curve of a specific radius does not actually give the optimal wire radius for a specific frequency.

To find an optimal wire radius, we maximize the quality factor $Q$ for a given plasma frequency $f_p$, shown by the dashed line in Fig.~\ref{fig:radius}. We see that the maximal $Q$ falls as $1/\sqrt{f_p}$ following the skin depth, assuming that the conductance is constant. Numerically, the maximum quality factor for a given frequency occurs when $r\simeq\lambda_c/11$. 
As this trade-off is caused by the geometry of the system, the optimal thickness of the wires is largely unaffected by the conductance of the wires, which only modifies the extrema of $Q$ through the Bessel functions in Eq.~\eqref{eq:Q1}. Note that as the wire thickness starts to become an appreciable fraction of $\lambda_c$, the assumption that they are one-dimensional objects breaks down, which would necessitate modifying the effective medium approach. Thus while these results are indicative of the $Q$ that can be achieved, a practical design should be fine-tuned through numerical simulation.

To give a sense of scale, Refs.~\cite{Lawson2019,Gelmini:2020kcu} assumed $Q=100$ over a range of $2-100\,{\rm GHz}$ in a cryogenic environment. However, as we have shown even with room temperature copper, it is possible to get $Q>10^3$ over this entire range, a gain of more than an order of magnitude. In a cryogenic environment we would expect further gains, ultimately being limited by the anomalous skin effect. For an example, a cavity designed specifically to search for axion dark matter was demonstrated to improve in quality factor by a factor of $\sim 3$ at 12\,GHz by moving to cryogenic temperatures~\cite{Ahn:2017smt}. Further gains could be made by moving to superconducting wires, though this may present design challenges in strong magnetic fields. We have also demonstrated that with some simple optimization one can noticeably improve the quality factor for a given frequency. This allows for high-quality WM resonators to be manufactured, with particularly interesting applications in the low-frequency regime. 
 
\section{Resonator eigenmode simulations}
While analytic formulas provide great simplicity and transparency, they will necessarily neglect the fine details of the system. Further, the region of validity of such a homogenized effective medium approach must be explored. Testing the validity of the analytic approach requires a comparison with both simulations and physical measurements. To accomplish the former, we use the Frequency Domain Solver in CST MWS 2020. 

\begin{figure*}[t]
\centering
    \begin{subfigure}[b]{.35\linewidth}
\includegraphics[width=\linewidth]{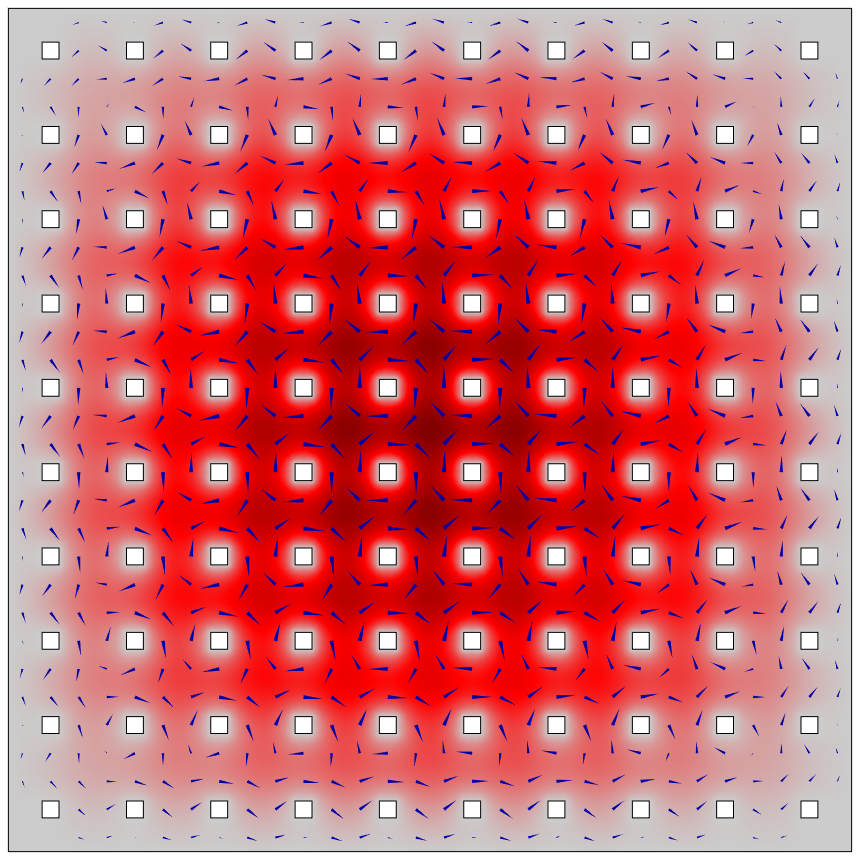}
\caption{}\label{fig:tm110}
\end{subfigure}
\begin{subfigure}[b]{.35\linewidth}
\includegraphics[width=\linewidth]{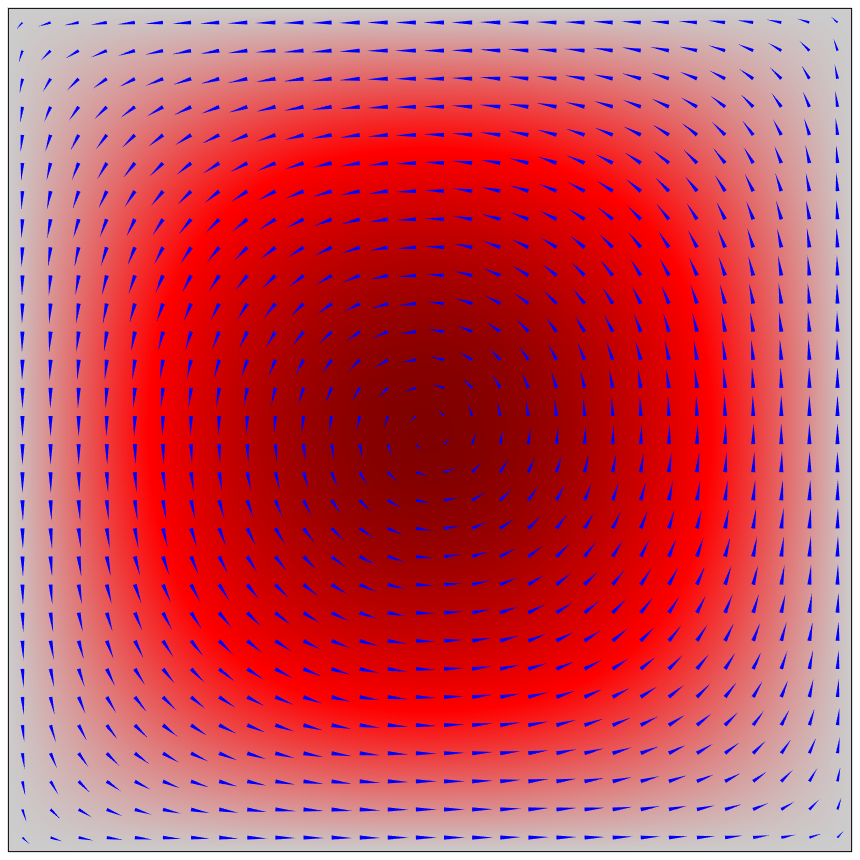}
\caption{}\label{fig:tm110h}
\end{subfigure}
\begin{subfigure}[b]{0.063\linewidth}
\includegraphics[width=\linewidth]{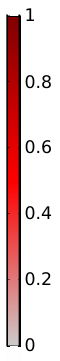}
\hspace{10pt}
\end{subfigure}

    \caption{The TM110 mode structure in the $xy$ cross section of a WM loaded cavity calculated in a CST MWS Eigenmode solver. The amplitudes of the $E$-field scaled via its maximum value ($E_{\rm max}$), $E/E_{\rm max}$, are shown via the colour map (red). It is clear that the distribution of the electric field is almost the same in two plots and corresponds to the one given by Eq.~\eqref{eq:fields}. The log-scaled in-plane components of the $H$-field are shown via the blue arrows. The direction (amplitude) of the $H$-field corresponds to the direction (magnitude) of the arrows.  {\em Left:} A full simulation with metal wires. The WM consists of a $10\times 10$ array of square wires of thickness $2r$, with spacing $a=1$\,cm and $r=1$\,mm inside a metal cavity as shown in Fig.~\ref{fig:wm_res_3d}. {\em Right:} A simulation with an effective medium using the analytic formula in Ref.~\cite{Belov2002} using the same parameters, except replacing the square wires of the numerical simulation with radius 1\,mm. 
    }
    \label{fig:xy_tm110}
\end{figure*}

To look at the mode structure directly, we will take the case of a $10 \times 10$ wire array, using $r=1$\,mm copper wires with a period of $a=1$\,mm (arranged as shown in Fig.~\ref{fig:wm_res_3d}). We compare the full wire treatment with a completely homogeneous effective medium, with properties as calculated in Sec.~\ref{analytic}. To start, we plot the $E$ and $H$-field distributions of the TM110 mode in Fig.~\ref{fig:xy_tm110}. Comparing the two red colour maps showing the value of the $E$-field we see that the presence of the individual wires does not have a significant impact on the overall mode structure for the $E$-field, except a local decrease in the immediate proximity of the wires.

However, the sharp decrease in $E$-field around the wires causes significant $H$-fields to be generated. Because of this the magnetic field circles around individual wires, and the highest amplitudes are achieved not at the centers of the walls, but at the cavity center (same as for the $E$-field). Indeed, it is the mutual inductance of the wires that is responsible for the collective oscillation of WM~\cite{Pendry:1998}. While this change does not have an impact for the unloaded $Q$, it may be significant when we consider the coupling of external circuits to the resonator. An empty cavity would  effectively couple to a magnetic probe inserted at the internal side of the wall. In a wire-filled cavity, however, the strong local $H$-fields near the wires may affect the loop antenna very differently. For instance, since the magnetic fields of any two neighbouring wires to an extent counteract each other at the point between them, placing a loop antenna there would lead to a significantly weaker signal. Note that the $H$-fields around the wires are approximately three times stronger than the strongest $H$-fields in the homogenized effective medium approach. 
Therefore SMA (sub-miniature A-version) ports with an 8-mm-long extended core placed at the point of the maximum of the $E$-field on the top and/or bottom walls were used as monopole antenna probes in both simulations and experiments.

\begin{figure}[h]
\centering
    \begin{subfigure}[b]{.49\linewidth}
\includegraphics[width=\linewidth]{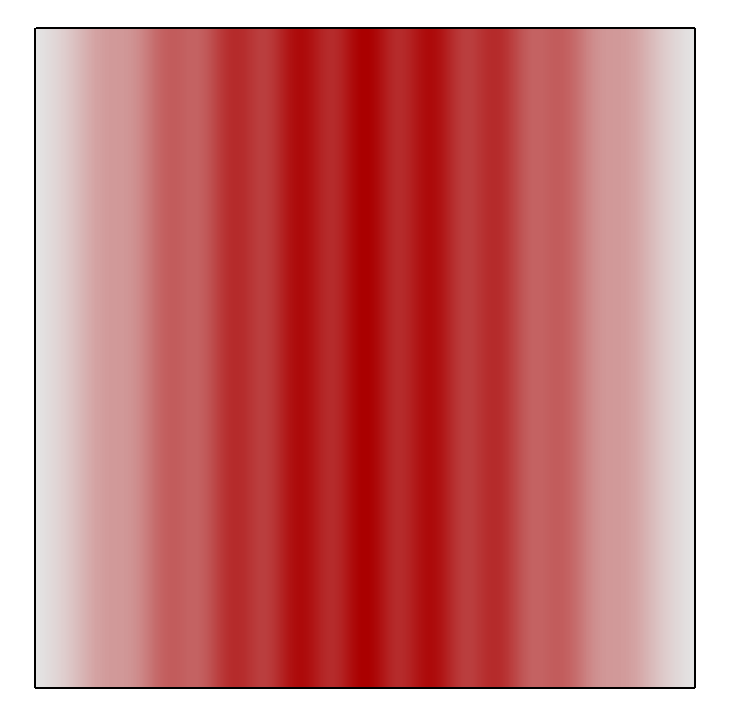}
\caption{}\label{fig:tm110no_ports}
\end{subfigure}
\begin{subfigure}[b]{.49\linewidth}
\includegraphics[width=\linewidth]{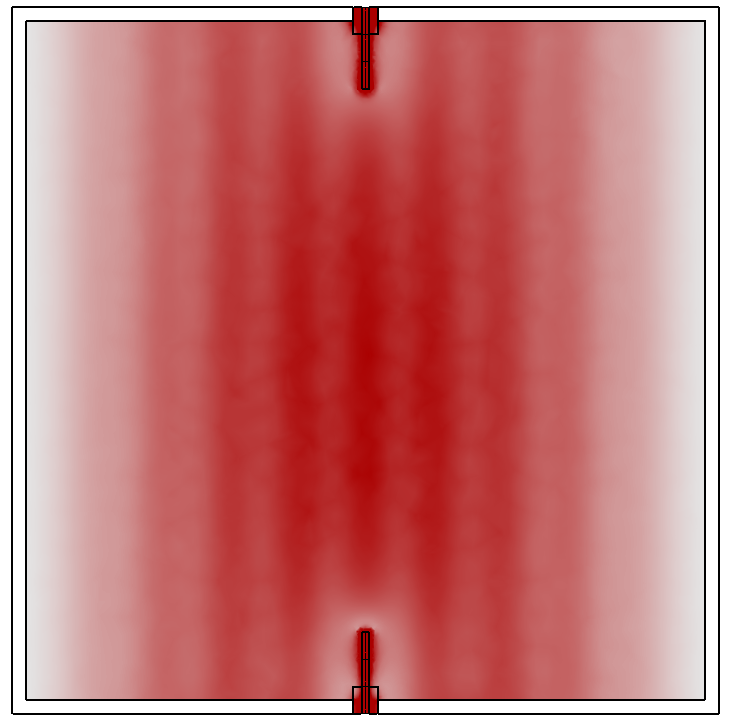}
\caption{}\label{fig:tm110ports}
\end{subfigure}
    \caption{Cross section of the $E$-field mode structure of the TM110 mode of a WM loaded resonator through the $x,z$-directions found with the CST MWS Eigenmode solver. The WM consists of a $10\times 10$ array of square wires of thickness $2r$, with spacing $a=1$\,cm and $r=1$\,mm inside a metal cavity as shown in Fig.~\ref{fig:wm_res_3d}. Darker red indicates a stronger $E$-field. {\em Left:} TM110 mode of an isolated resonator (i.e., no antenna).  {\em Right:} TM110 mode of the resonator with two SMA ports acting as 8-mm-long monopole probes added at the center of the top and bottom walls. The configuration mirrors the one used in the experiment.}
    \label{fig:xz_tm110_compar}
\end{figure}

In order to couple to the resonant cavity, two SMA ports were added at the centers of the walls at which the wires were terminated. Being placed at the point of the maximum electric field at these walls, the SMA probes are very strongly coupled to the mode field making the resonator heavily loaded by these lumped ports. 
The vertical position of the probes allows us to only excite the TM modes of the resonator, without coupling to the numerous TE modes of the cavity, which are weakly influenced by the wires. As can be seen from the comparison in Fig. \ref{fig:xz_tm110_compar}, in terms of field magnitude the presence of probes changes the field structure in their vicinity, but does little in terms of affecting the mode structure at large. However, a slight non-homogeneity is added in terms of the phase of the dominating $E_z$ component.

Thus we can see that while the effective medium approach allows for a good prediction for the overall properties of a WM resonator, the presence of wires leads to local distortions which may have an impact on the coupling of antennas to the system. 

\section{Experiment}
\begin{figure}[h]
\centering
    \begin{subfigure}[b]{.49\linewidth}
\includegraphics[width=\linewidth]{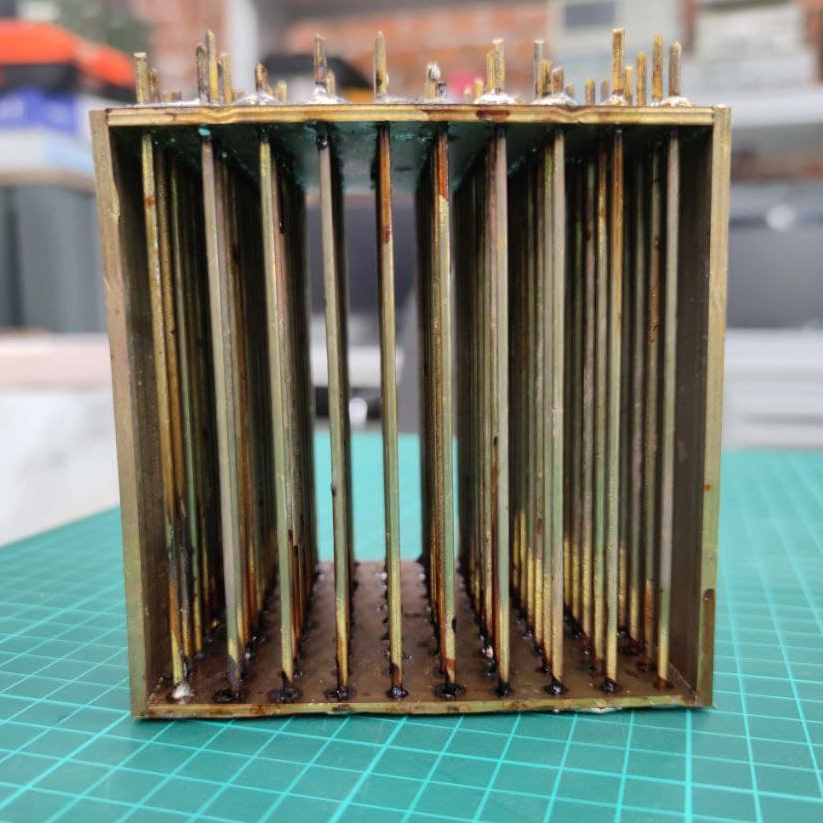}
\caption{}\label{fig:itmo_exp_1}
\end{subfigure}
\begin{subfigure}[b]{.49\linewidth}
\includegraphics[width=\linewidth]{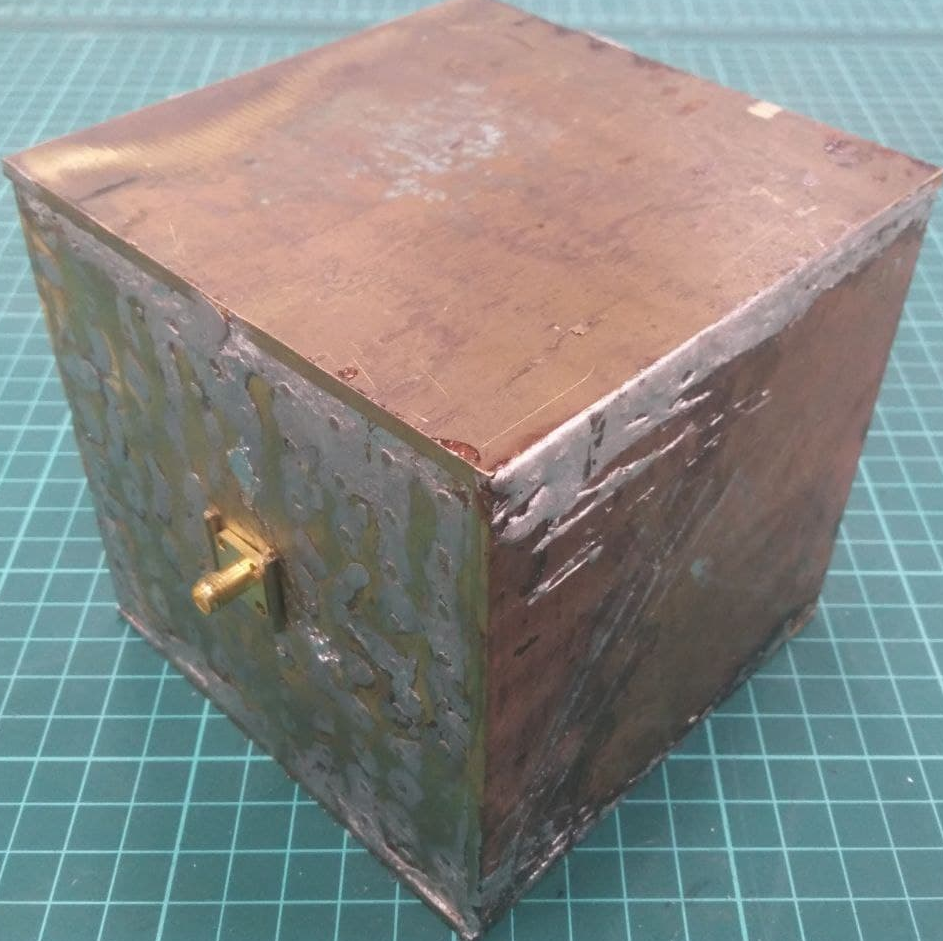}
\caption{}\label{fig:itmo_exp_2}
\end{subfigure}
    \caption{Photos of the experimental prototype WM resonator. The resonator is a 10$\times$10 array of radius $1$\,mm circular cross-section brass wires placed with a period $a=1$\,cm. The wires are inserted into the holes in the brass walls and soldered to ensure an electric connection. Two 8-mm-long SMA connectors acting as monopole antenna probes are inserted at the centers of the walls to which the wires are connected. {\em Left:} The wires inside the resonator in an unfinished state. {\em Right:} The finished resonator fully enclosed in a cavity.  }
    \label{fig:itmo_exp}
\end{figure}

Lastly, analytic and numerical solutions must always be validated by actual physical measurements. To this end,  
a brass prototype was built with the aim of testing whether a $Q$-factor on the order of thousands could be achieved at room temperature by enclosing the wires within a cavity. Yellow brass wires (65\% copper, 35\% zinc) of $r=1$\,mm  were placed and welded within a $10\times 10\times 10$\,cm brass cube with a period $a=1$\,cm. The lateral walls were spaced $a/2=0.5$\,cm away from the WM sample. To facilitate measurements, two SMA ports were attached to the top and bottom walls to which the wires were welded, as shown in Fig.~\ref{fig:itmo_exp}. The two ports acted as 8-mm long monopole probes coupled to the TM modes of the resonator. The center contact of the ports had a diameter of 1.3~mm and was made of beryllium copper. The insulator dielectric had a diameter of 4.1~mm and was made of polytetrafluoroethylene (PTFE). The conductivity of the metal used in our experiment was measured in the DC regime to be $1.51\times 10^7$~S/m. This conductivity is only slightly lower than the tabulated value $1.59\times 10^7$~S/m of the 65\% brass that was used in the CST model for the numerical comparisons.  

To measure the system, we used an Agilent E8362C vector network analyzer (VNA) in the range of 11-13.5 GHz connected to SMA connectors acting as monopole antennas. 
We first measured the coupling coefficient of each of the ports by attaching the other port to a matched load and measuring the reflection coefficient $S_{11}$ at the port in question. After employing curve fitting to the resulting $S$-parameters to account for the coupling losses at the ports, as described in Ref.~\cite{Darko2005}, we obtained a pair of coupling coefficients $\kappa_{m1}, \kappa_{m2}$ for the matched setup. 

We can convert the coupling coefficients of a matched cavity to input and output coupling coefficients $\kappa_1,\kappa_2$ via~\cite{Darko2005} 
\begin{subequations}
\begin{align}
    \kappa_1 &= \kappa_{m1} \frac{1 + \kappa_{m2}}{1 - \kappa_{m1}\kappa_{m2}}\,, \\
    \kappa_2 &= \kappa_{m2} \frac{1 + \kappa_{m1}}{1 - \kappa_{m1}\kappa_{m2}}\,.
\end{align}
\end{subequations}
The overall coupling coefficient can then be found as the sum of the input and output ones,
\begin{equation}
    \kappa = \kappa_1 + \kappa_2\,.
\end{equation}

\begin{figure}[t]
\centering
    \includegraphics[width=\linewidth]{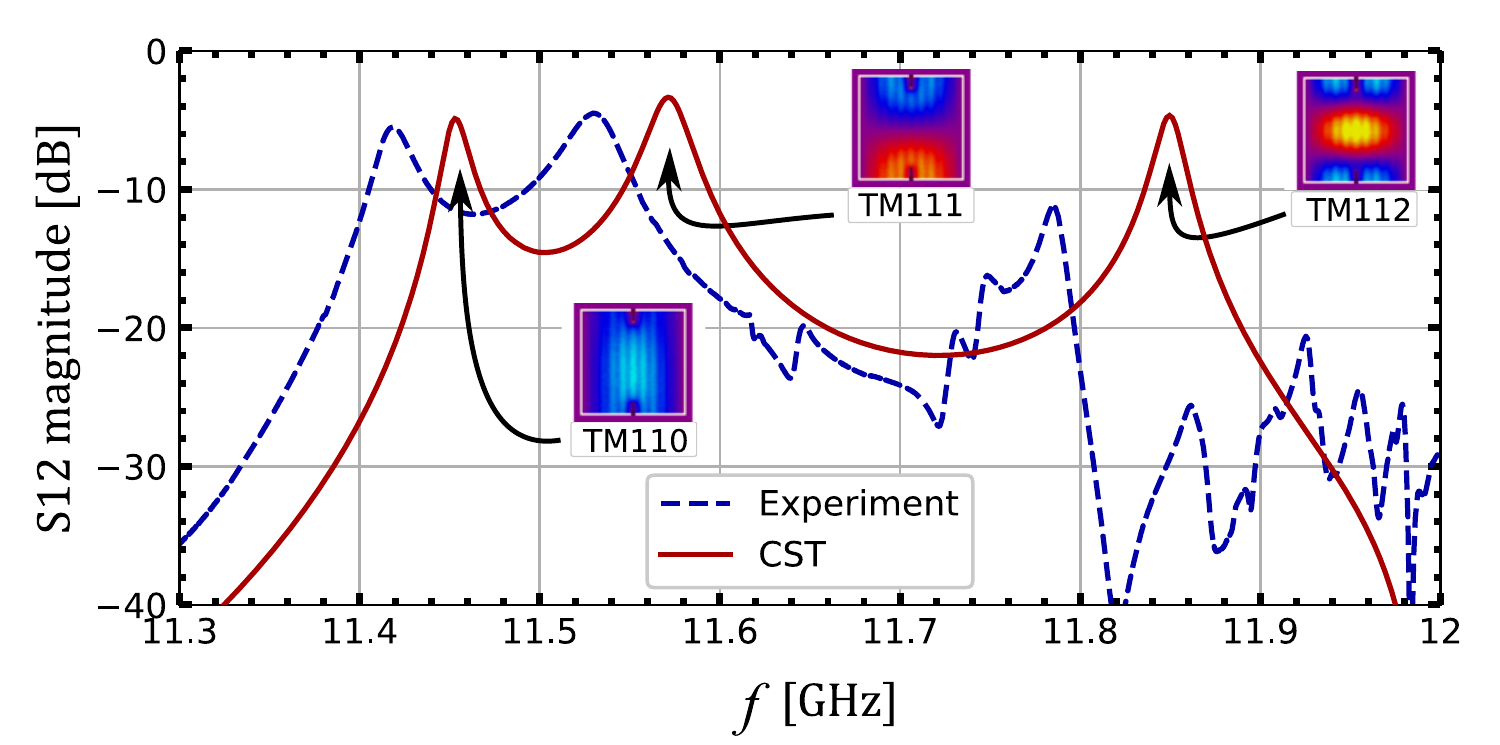}
    \caption{S12 (transmission) parameters for a WM consisting of a $10\times 10$ wire array with spacing $a=1$\,cm inside a metal cavity. We depict both a CST simulation (red line) similar to that shown in Fig.~\ref{fig:wm_res_3d} and the measured values of the prototype experiment (blue dashed line) as shown in Fig~\ref{fig:itmo_exp}. The CST simulation assumes 65\% brass wires with radius $1$\,mm, whereas the experiment uses wires with the same radius but has a measured DC conductivity of $1.51\times 10^7$\,S/m (slightly lower than 65\% brass). Circular cross-section wires are used in both the experiment and the simulation. The $xz$ cross-sections of the corresponding TM modes are shown as insets next to the resonances in the CST simulation.}
    \label{fig:s12}
\end{figure}

Thus armed with the coupling coefficient, we used a transmission-type measurement to study the $Q$-factor of the system. In Fig.~\ref{fig:s12} we compare the numerical simulation with the S12 measurement. Apart from the lowest TM110 mode around 11.45~GHz, two other modes - TM111 and TM112 can also be seen at higher frequencies.
While there is a good overall agreement, there is a discrepancy of about 33~MHz in the frequency of the resonances. This discrepancy is likely caused by the finite accuracy available for a given mesh and can likely be neutralized by increasing its density.
The loaded $Q$-factor $Q_L$, corresponding to the full width at half maximum, can be read off from the 3-dB width of the $S_{12}$ maximum corresponding to the mode in question. To translate this reading into an unloaded $Q$-factor $Q_U$ we can use the coupling coefficient via
\begin{equation}
    Q_U = Q_L (1 + \kappa)\,.
\end{equation}

We summarize the key parameters ($\kappa,Q,f_{\rm res}$, etc.) in Table~\ref{table:itmo-proto}. The most significant difference occurs in the unloaded quality factor $Q_U$: the experimental prototype is noticeably more lossy than the simulated one, that results in wider bandwidth and lower $Q$. The difference in the coupling coefficients might be attributed to the non-ideal electrical connection of the port and leakage of energy in the experimental case resulting in lower coupling.

\begin{figure}[t]
\centering
    \includegraphics[width=1\linewidth]{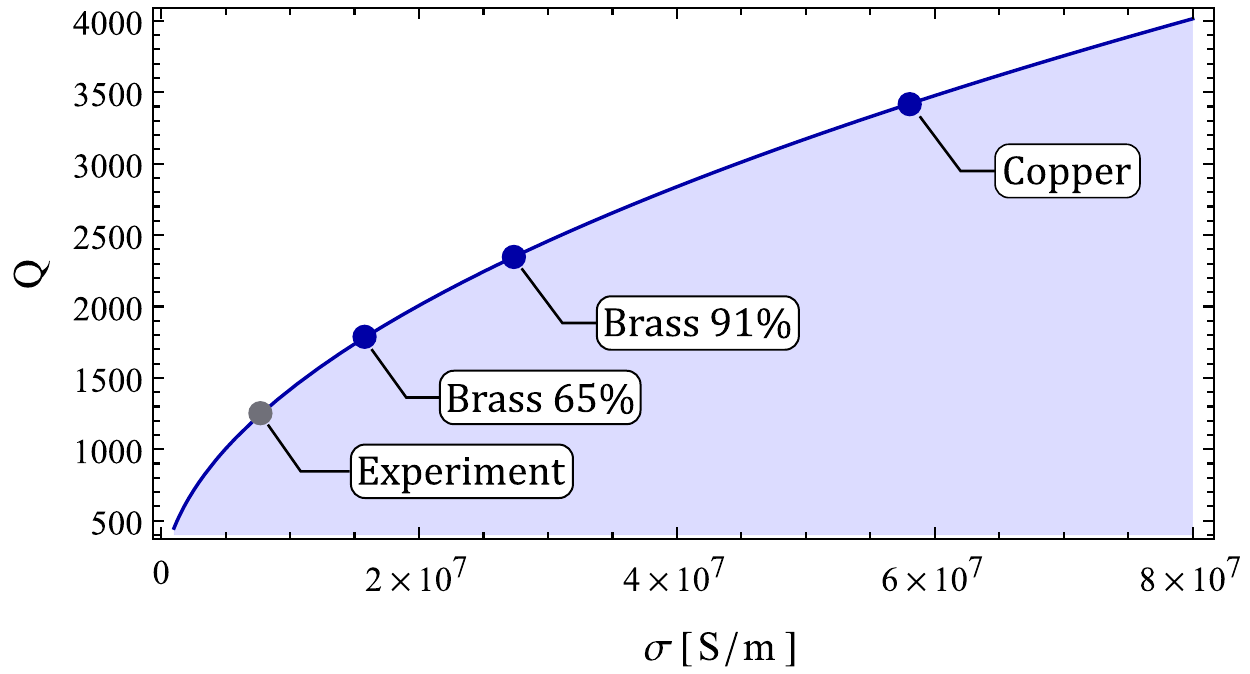}
    \caption{Dependence of unloaded quality factor $Q$ on the conductivity of the wires $\sigma$ for an infinite wire media. The other parameters of the resonator are the same as for Fig.~\ref{fig:wm_res_Q}. The three blue points mark maximum unloaded quality factors achievable with wires made of common metals, 65\% brass, 91\% brass, and pure copper. The gray point shows the quality factor observed in the experiment and an equivalent conductivity that would result in the same value for the otherwise ideal conditions.
    }
    \label{fig:sigma}
\end{figure}

The extra losses in the experimental setup may be attributed to a number of factors: the presence of an oxidized layer on the wires (while it does not influence the DC conductivity averaged over the wire cross section, the skin effect means that the majority of the current occurs in the periphery of the wire), the sub-millimeter roughness of the wires surface, the presence of the drops of solder inside the resonator, and possible misalignment of the wires. 
Using the expressions from our analytical model, these various losses can be all factored into an effective conductivity $\sigma_{\text{eff}}$, 
which is decreased compared to the measured one (already lower than the tabulated one) and turns out to be equal to $0.77\times 10^7$~S/m, i.e., about twice lower than the actual conductivity of brass. In Fig.~\ref{fig:sigma} we present the dependence of the unloaded quality factor on the conductivity $\sigma$ calculated analytically for several metals in comparison with the $\sigma_{\text{eff}}$ observed in the experiment to better visualise the discrepancy and the extent to which the quality factor may be improved with more refined prototype and measurement.

 \begin{table}[t]
\centering
\begin{tabular}{@{}lcc@{}}

\textbf{}                                 & \multicolumn{1}{l}{\textbf{Experimental}} & \multicolumn{1}{l}{\textbf{Numerical}} \\ \midrule
\textbf{Frequency {[}GHz{]}}              & 11.420                             & 11.453                               \\
\textbf{Bandwidth {[}GHz{]}}              & 0.022                              & 0.016                                 \\
\textbf{Loaded Q}                         & 509                                & 735                                \\
\textbf{Coupling coefficient}             & 1.34                               & 1.82                                \\
\textbf{Unloaded Q}                       & 1194                               & 2074                               \\ \bottomrule
\end{tabular}

\caption{Comparison between numerical simulations in CST and measurements of the experimental prototype for the key parameters of the TM110 mode. The WM consists of a $10\times 10$ wire array with spacing $a=1$\,cm inside a metal cavity. The CST simulation assumes 65\% brass wires with radius $1$\,mm, whereas the experiment uses wires with the same radius but has a measured DC conductivity of $1.51\times 10^7$\,S/m (slightly lower than 65\% brass). Circular cross-section wires are used in both the experiment and the simulation.}
\label{table:itmo-proto}
\end{table}

While the resulting value of the unloaded Q-factor is 40\% lower than expected, the value obtained still speaks in favour of the quality factors on the order of thousands being achievable with high-quality copper wires (and manufacturing). Further investigations explore the sources of increased losses to allow for the highest possible quality factors.

\section{Conclusions}
In this work we have analyzed the behavior of a cubic microwave resonator filled with uniaxial wire medium analytically, numerically, and experimentally. One particularly interesting use of such a system is to build a plasma haloscope~\cite{Lawson2019}; however, previous work made no computation of the quality factor of such a system, and assumed a homogenized effective medium model. When a WM is enclosed inside a metal cavity, we find that the primary source of loss comes from the conductivity of the wires themselves, with wall losses only playing a role for small systems. While the quality factor decreases as the size of the system increases, it plateaus at the value expected from wire losses alone. Thus the system size can be increased to allow for large volume (and thus high signal power) haloscopes without negatively impacting the resulting quality factors. We also discussed the limits in which the system can be treated as a single-mode resonator: when the system size is larger than $\sim\sqrt{Q}\lambda_c$ such a description breaks down. While a plasma haloscope can operate in such a limit, the decay length in media would require a multi-antenna readout design (such as a phase-matched array).

We performed numerical simulations in CST, which showed a very strong agreement with the analytic model. This indicates that additional boundary conditions are not needed for the case of a wire medium with an $a/2$ spacing between the wires and the walls, which validates the homogenization approach. These numerical simulations were then compared to an experimental prototype, which showed good agreement in the mode structure and loaded quality factors. However, the resulting unloaded quality factor was 40\% lower than expected, most likely due to flaws in the manufacturing of the prototype. Promisingly, even with room temperature brass, a quality factor larger than a thousand was easily obtained, more than an order of magnitude higher than assumed in Ref.~\cite{Lawson2019}. With high quality copper, machining and moving to a cryogenic environment we expect that as much as an additional order of magnitude can be gained.

We have studied in detail for the first time the behaviour and quality factor of WM loaded resonators, finding that they are extremely promising for the purpose of detecting axion and dark photon dark matter. 

\section*{Acknowledgements}
The authors thank Maxim Gorlach, Jón Gudmundsson, Grigorij Karsakov, Tove Klaesson, Eugine Koreshin,  Mathew Lawson, Ivan Matchenya, and Karl van Bibber for helpful discussions and also thank the members of the ALPHA Consortium for discussion and support. R.B. thanks Gréta Horváthová for her continued support.
 A.J.M. is supported by the European Research
Council under Grant No. 742104 and by the Swedish Research Council (VR) under Dnr
2019-02337 “Detecting Axion Dark Matter In The Sky And In The Lab (AxionDM)”.
R.B. and P.B. were supported by 
a grant for scientific school \foreignlanguage{russian}{НШ}-2359.2022.4
and 
Priority 2030 Federal Academic Leadership Program.

\bibliographystyle{bibi}

\begin{thebibliography}{10}
	
	\bibitem{Pendry:1998}
	J.~B. Pendry, A.~J. Holden, D.~J. Robbins and W.~J. Stewart, \emph{{Low
			Frequency Plasmons in Thin-Wire Structures}},
	\href{https://doi.org/10.1088/0953-8984/10/22/007}{\emph{J. Phys. Condens.
			Matter} {\bfseries 10} 4785} (1998).
	
	\bibitem{Sim}
	C.~Simovski, P.~Belov, A.~Atraschenko and Y.~Kivshar, \emph{{Wire
			metamaterials: Physics and Applications}}, {\emph{Advanced Materials}
		{\bfseries 24} 4229} (2012).
	
	\bibitem{Sim0}
	V.~Milichko, A.~Shalin, I.~Mukhin, A.~Kovrov, A.~Krasilin, A.~Vinogradov,
	P.~Belov and C.~Simovski, \emph{{Solar Photovoltaics: Current State and
			Trends}}, {\emph{Physics Uspekhi} {\bfseries 186} 727} (2016).
	
	\bibitem{Forati2015}
	E.~Forati, G.~W. Hanson and D.~{F. Sievenpiper}, \emph{{An Epsilon-Near-Zero
			Total-Internal-Reflection Metamaterial Antenna}},
	\href{https://doi.org/10.1109/TAP.2015.2405559}{\emph{IEEE Trans. Antennas
			Propag.} {\bfseries 63} 1909} (2015).
	
	\bibitem{Si}
	C.~Simovski, S.~Maslovski, I.~Nefedov, S.~Kosulnikov, P.~Belov and
	S.~Tretyakov, \emph{{Hyperlens Makes Thermal Emission Strongly
			Super-Planckian}}, {\emph{Photonics and Nanostructures Fundamentals and
			Applications} {\bfseries 13} 31} (2015).
	
	\bibitem{Li}
	Y.~Li, J.~Lv, Q.~Gu, S.~Hu, Z.~Li, X.~Jiang, Y.~Ying and G.~Si,
	\emph{{Metadevices with Potential Practical Applications}},
	\href{https://doi.org/10.3390/molecules24142651}{\emph{Molecules} {\bfseries
			24} 2651} (2019).
	
	\bibitem{Sim1}
	M.~Mirmoosa, S.~Kosulnikov and C.~Simovski, \emph{{Magnetic Hyperbolic
			Metamaterial of High-Index Nanowires}}, {\emph{Physical Review B} {\bfseries
			94} 075138} (2016).
	
	\bibitem{Sim2}
	M.~Mirmoosa, S.~Kosulnikov and C.~Simovski, \emph{{Double Resonant Wideband
			Purcell Effect in Wire Metamaterials}}, {\emph{Journal of Optics} {\bfseries
			18} 095101} (2016).
	
	\bibitem{Lawson2019}
	M.~Lawson, A.~J. Millar, M.~Pancaldi, E.~Vitagliano and F.~Wilczek,
	\emph{{Tunable Axion Plasma Haloscopes}},
	\href{https://doi.org/10.1103/PhysRevLett.123.141802}{\emph{Phys. Rev. Lett.}
		{\bfseries 123} 141802} (2019)
	[\href{https://arxiv.org/abs/1904.11872}{{\ttfamily 1904.11872}}].
	
	\bibitem{Silveirinha2006}
	M.~Silveirinha and N.~Engheta, \emph{{{Tunneling of Electromagnetic Energy
				through Subwavelength Channels and Bends Using $\epsilon$-Near-Zero
				Materials}}},
	\href{https://doi.org/10.1103/PhysRevLett.97.157403}{\emph{Phys. Rev. Lett.}
		{\bfseries 97} 157403} (2006).
	
	\bibitem{Alu2007}
	A.~Al{\`{u}}, M.~G. Silveirinha, A.~Salandrino and N.~Engheta,
	\emph{{Epsilon-Near-Zero Metamaterials and Electromagnetic Sources: Tailoring
			the Radiation Phase Pattern}},
	\href{https://doi.org/10.1103/PhysRevB.75.155410}{\emph{Phys. Rev. B -
			Condens. Matter Mater. Phys.} {\bfseries 75} 155410} (2007).
	
	\bibitem{Zhou2010}
	R.~Zhou, H.~Zhang and H.~Xin, \emph{{Metallic Wire Array as Low-Effective Index
			of Refraction Medium for Directive Antenna}},
	\href{https://doi.org/10.1109/TAP.2009.2036282}{\emph{IEEE Trans. Antennas
			Propag.} {\bfseries 58} 79} (2010).
	
	\bibitem{Sim3}
	M.~Mirmoosa, S.~Kosulnikov and C.~Simovski, \emph{{Unbounded Spatial Spectrum
			of Propagating Waves in a Polaritonic Wire Medium}}, {\emph{Physical Review
			B} {\bfseries 92} 075139} (2015).
	
	\bibitem{Peccei:1977hh}
	R.~D. Peccei and H.~R. Quinn, \emph{{CP Conservation in the Presence of
			Instantons}}, \href{https://doi.org/10.1103/PhysRevLett.38.1440}{\emph{Phys.
			Rev. Lett.} {\bfseries 38} 1440} (1977).
	
	\bibitem{Weinberg:1977ma}
	S.~Weinberg, \emph{{A New Light Boson?}},
	\href{https://doi.org/10.1103/PhysRevLett.40.223}{\emph{Phys. Rev. Lett.}
		{\bfseries 40} 223} (1978).
	
	\bibitem{Wilczek:1977pj}
	F.~Wilczek, \emph{{Problem of Strong P and T Invariance in the Presence of
			Instantons}}, \href{https://doi.org/10.1103/PhysRevLett.40.279}{\emph{Phys.
			Rev. Lett.} {\bfseries 40} 279} (1978).
	
	\bibitem{Preskill:1982cy}
	J.~Preskill, M.~B. Wise and F.~Wilczek, \emph{{Cosmology of the Invisible
			Axion}}, \href{https://doi.org/10.1016/0370-2693(83)90637-8}{\emph{Phys.
			Lett. B} {\bfseries 120} 127} (1983).
	
	\bibitem{Abbott:1982af}
	L.~Abbott and P.~Sikivie, \emph{{A Cosmological Bound on the Invisible Axion}},
	\href{https://doi.org/10.1016/0370-2693(83)90638-X}{\emph{Phys. Lett. B}
		{\bfseries 120} 133} (1983).
	
	\bibitem{Dine:1982ah}
	M.~Dine and W.~Fischler, \emph{{The Not So Harmless Axion}},
	\href{https://doi.org/10.1016/0370-2693(83)90639-1}{\emph{Phys. Lett. B}
		{\bfseries 120} 137} (1983).
	
	\bibitem{Bergstrom:2000pn}
	L.~Bergström, \emph{{Nonbaryonic Dark Matter: Observational Evidence and
			Detection Methods}},
	\href{https://doi.org/10.1088/0034-4885/63/5/2r3}{\emph{Rept. Prog. Phys.}
		{\bfseries 63} 793} (2000)
	[\href{https://arxiv.org/abs/hep-ph/0002126}{{\ttfamily hep-ph/0002126}}].
	
	\bibitem{Jaeckel:2010ni}
	J.~Jaeckel and A.~Ringwald, \emph{{The Low-Energy Frontier of Particle
			Physics}},
	\href{https://doi.org/10.1146/annurev.nucl.012809.104433}{\emph{Ann. Rev.
			Nucl. Part. Sci.} {\bfseries 60} 405} (2010)
	[\href{https://arxiv.org/abs/1002.0329}{{\ttfamily 1002.0329}}].
	
	\bibitem{Feng:2010gw}
	J.~L. Feng, \emph{{Dark Matter Candidates from Particle Physics and Methods of
			Detection}},
	\href{https://doi.org/10.1146/annurev-astro-082708-101659}{\emph{Ann. Rev.
			Astron. Astrophys.} {\bfseries 48} 495} (2010)
	[\href{https://arxiv.org/abs/1003.0904}{{\ttfamily 1003.0904}}].
	
	\bibitem{Sikivie:1983ip}
	P.~Sikivie, \emph{{Experimental Tests of the Invisible Axion}},
	\href{https://doi.org/10.1103/PhysRevLett.51.1415}{\emph{Phys. Rev. Lett.}
		{\bfseries 51} 1415} (1983). [Erratum: Phys.Rev.Lett. 52, 695 (1984)].
	
	\bibitem{Rybka:2014xca}
	{\scshape ADMX} Collaboration, G.~Rybka, \emph{{Direct Detection Searches for
			Axion Dark Matter}},  in \emph{{Proceedings, 13th international conference on
			Topics in Astroparticle and Underground Physics (TAUP 2013): Asilomar,
			California, September 8-13, 2013}}, Elsevier BV, 2014,
	\href{https://doi.org/10.1016/j.dark.2014.05.003}{DOI}.
	
	\bibitem{Woohyun:2016}
	W.~Chung, \emph{{Launching Axion Experiment at CAPP/IBS in Korea}},  in
	\emph{{Proceedings, 12th Patras workshop on axions, WIMPs and WISPs: Jeju
			Island, South Korea, June 20-24, 2016}}, DESY: Hamburg, Germany (2017)
	30--34, \href{https://doi.org/10.3204/DESY-PROC-2009-03/Chung_Woohyun}{DOI}.
	
	\bibitem{Buschmann:2021sdq}
	M.~Buschmann, J.~W. Foster, A.~Hook, A.~Peterson, D.~E. Willcox, W.~Zhang and
	B.~R. Safdi, \emph{Dark matter from axion strings with adaptive mesh
		refinement}, \href{https://doi.org/10.1038/S41467-022-28669-Y}{\emph{Nature
			Communications} {\bfseries 13} } (2022)
	[\href{https://arxiv.org/abs/2108.05368}{{\ttfamily 2108.05368}}].
	
	\bibitem{Goryachev:2017wpw}
	M.~Goryachev, B.~T. Mcallister and M.~E. Tobar, \emph{{Axion Detection with
			Negatively Coupled Cavity Arrays}},
	\href{https://doi.org/10.1016/j.physleta.2017.09.016}{\emph{Phys. Lett. A}
		{\bfseries 382} 2199} (2018)
	[\href{https://arxiv.org/abs/1703.07207}{{\ttfamily 1703.07207}}].
	
	\bibitem{Melcon:2018dba}
	A.~{\'{A}}. Melc{\'{o}}n, S.~A. Cuendis, C.~Cogollos, A.~D{\'{\i}}az-Morcillo,
	B.~Döbrich, J.~D. Gallego, B.~Gimeno, I.~G. Irastorza, A.~J.
	Lozano-Guerrero, C.~Malbrunot et~al., \emph{Axion searches with microwave
		filters: the {RADES} project},
	\href{https://doi.org/10.1088/1475-7516/2018/05/040}{\emph{Journal of
			Cosmology and Astroparticle Physics} {\bfseries 2018} 040} (2018).
	
	\bibitem{Melcon:2020xvj}
	A.~{\'{A}}. Melc{\'{o}}n, S.~A. Cuendis, C.~Cogollos, A.~D{\'{\i}}az-Morcillo,
	B.~Döbrich, J.~D. Gallego, J.~M.~G. Barceló, B.~Gimeno, J.~Golm, I.~G.
	Irastorza et~al., \emph{{Scalable Haloscopes for Axion Dark Matter Detection
			in the 30$\mu$eV range with RADES}},
	\href{https://doi.org/10.1007/JHEP07(2020)084}{\emph{JHEP} {\bfseries 2020}
		084} (2020) [\href{https://arxiv.org/abs/2002.07639}{{\ttfamily
			2002.07639}}].
	
	\bibitem{Jeong:2020cwz}
	J.~Jeong, S.~Youn, S.~Bae, J.~Kim, T.~Seong, J.~E. Kim and Y.~K. Semertzidis,
	\emph{{Search for Invisible Axion Dark Matter with a Multiple-Cell
			Haloscope}},
	\href{https://doi.org/10.1103/PhysRevLett.125.221302}{\emph{Phys. Rev. Lett.}
		{\bfseries 125} 221302} (2020)
	[\href{https://arxiv.org/abs/2008.10141}{{\ttfamily 2008.10141}}].
	
	\bibitem{Carosi:2020akt}
	G.~Carosi, R.~Cervantes, S.~Kimes, P.~Mohapatra, R.~Ottens and G.~Rybka,
	\emph{{Orpheus: Extending the ADMX QCD Dark-Matter Axion Search to Higher
			Masses}}, \href{https://doi.org/10.1007/978-3-030-43761-9_20}{\emph{Springer
			Proc. Phys.} {\bfseries 245} 169} (2020).
	
	\bibitem{Quiskamp:2020yrx}
	A.~P. Quiskamp, B.~T. McAllister, G.~Rybka and M.~E. Tobar,
	\emph{{Dielectric-Boosted Sensitivity to Cylindrical Azimuthally Varying
			Transverse-Magnetic Resonant Modes in an Axion Haloscope}},
	\href{https://doi.org/10.1103/PhysRevApplied.14.044051}{\emph{Phys. Rev.
			Applied} {\bfseries 14} 044051} (2020)
	[\href{https://arxiv.org/abs/2006.05641}{{\ttfamily 2006.05641}}].
	
	\bibitem{Kim2020}
	J.~Kim, S.~Youn, J.~Jeong, W.~Chung, O.~Kwon and Y.~K. Semertzidis,
	\emph{{Exploiting Higher-Order Resonant modes for Axion Haloscopes}},
	\href{https://doi.org/10.1088/1361-6471/ab5ace}{\emph{J. Phys. G Nucl. Part.
			Phys.} {\bfseries 47} } (2020)
	[\href{https://arxiv.org/abs/1910.00793}{{\ttfamily 1910.00793}}].
	
	\bibitem{Alesini2020}
	D.~Alesini, C.~Braggio, G.~Carugno, N.~Crescini, D.~D'Agostino, D.~{Di
		Gioacchino}, R.~{Di Vora}, P.~Falferi, U.~Gambardella, C.~Gatti et~al.,
	\emph{{High Quality Factor Photonic Cavity for Dark Matter Axion Searches}},
	\href{https://doi.org/10.1063/5.0003878}{\emph{Rev. Sci. Instrum.} {\bfseries
			91} } (2020) [\href{https://arxiv.org/abs/2002.01816}{{\ttfamily
			2002.01816}}].
	
	\bibitem{Alesini2021}
	D.~Alesini, C.~Braggio, G.~Carugno, N.~Crescini, D.~{D' Agostino}, D.~{Di
		Gioacchino}, R.~{Di Vora}, P.~Falferi, U.~Gambardella, C.~Gatti et~al.,
	\emph{{Realization of a High Quality Factor Resonator with Hollow Dielectric
			Cylinders for Axion Searches}},
	\href{https://doi.org/10.1016/j.nima.2020.164641}{\emph{Nucl. Instruments
			Methods Phys. Res. Sect. A Accel. Spectrometers, Detect. Assoc. Equip.}
		{\bfseries 985} 164641} (2021)
	[\href{https://arxiv.org/abs/2004.02754}{{\ttfamily 2004.02754}}].
	
	\bibitem{Horns:2012jf}
	D.~Horns, J.~Jaeckel, A.~Lindner, A.~Lobanov, J.~Redondo and A.~Ringwald,
	\emph{{Searching for WISPy Cold Dark Matter with a Dish Antenna}},
	\href{https://doi.org/10.1088/1475-7516/2013/04/016}{\emph{JCAP} {\bfseries
			2013} 016} (2013) [\href{https://arxiv.org/abs/1212.2970}{{\ttfamily
			1212.2970}}].
	
	\bibitem{Jaeckel:2013sqa}
	J.~Jaeckel and J.~Redondo, \emph{{An Antenna for Directional Detection of WISPy
			Dark Matter}},
	\href{https://doi.org/10.1088/1475-7516/2013/11/016}{\emph{JCAP} {\bfseries
			2013} 016} (2013) [\href{https://arxiv.org/abs/1307.7181}{{\ttfamily
			1307.7181}}].
	
	\bibitem{Suzuki:2015sza}
	J.~Suzuki, T.~Horie, Y.~Inoue and M.~Minowa, \emph{{Experimental Search for
			Hidden Photon CDM in the eV mass range with a Dish Antenna}},
	\href{https://doi.org/10.1088/1475-7516/2015/09/042}{\emph{JCAP} {\bfseries
			2015} 042} (2015) [\href{https://arxiv.org/abs/1504.00118}{{\ttfamily
			1504.00118}}].
	
	\bibitem{Experiment:2017icw}
	{\scshape FUNK Experiment} Collaboration, D.~Veberič, A.~Andrianavalomahefa,
	K.~Daumille, B.~Döbrich(CERN), R.~Engel, J.~Jaecke, M.~Kowalski, H.~J.~M.
	A.~Lindner(DESY, Zeuthen), J.~Redondo, M.~Roth et~al., \emph{{Search for
			Hidden-Photon Dark Matter with the FUNK Experiment}},
	\href{https://doi.org/10.22323/1.301.0880}{\emph{PoS} {\bfseries ICRC2017}
		880} (2018) [\href{https://arxiv.org/abs/1711.02958}{{\ttfamily
			1711.02958}}].
	
	\bibitem{BREAD:2021tpx}
	{\scshape BREAD} Collaboration, J.~Liu, K.~Dona, G.~Hoshino, S.~Knirck,
	N.~Kurinsky, M.~Malaker, D.~W. Miller, A.~Sonnenschein, M.~H. Awida, P.~S.
	Barry et~al., \emph{{Broadband Solenoidal Haloscope for Terahertz Axion
			Detection}},
	\href{https://doi.org/10.1103/PhysRevLett.128.131801}{\emph{Phys. Rev. Lett.}
		{\bfseries 128} 131801} (2022)
	[\href{https://arxiv.org/abs/2111.12103}{{\ttfamily 2111.12103}}].
	
	\bibitem{TheMADMAXWorkingGroup:2016hpc}
	{\scshape MADMAX Working Group} Collaboration, A.~Caldwell, G.~Dvali,
	B.~Majorovits, A.~Millar, G.~Raffelt, J.~Redondo, O.~Reimann, F.~Simon and
	F.~Steffen, \emph{{Dielectric Haloscopes: A New Way to Detect Axion Dark
			Matter}}, \href{https://doi.org/10.1103/PhysRevLett.118.091801}{\emph{Phys.
			Rev. Lett.} {\bfseries 118} 091801} (2017)
	[\href{https://arxiv.org/abs/1611.05865}{{\ttfamily 1611.05865}}].
	
	\bibitem{Baryakhtar:2018doz}
	M.~Baryakhtar, J.~Huang and R.~Lasenby, \emph{{Axion and Hidden Photon Dark
			Matter Detection with Multilayer Optical Haloscopes}},
	\href{https://doi.org/10.1103/PhysRevD.98.035006}{\emph{Phys. Rev. D}
		{\bfseries 98} 035006} (2018)
	[\href{https://arxiv.org/abs/1803.11455}{{\ttfamily 1803.11455}}].
	
	\bibitem{Chiles:2021gxk}
	J.~Chiles, I.~Charaev, R.~Lasenby, M.~Baryakhtar, J.~Huang, A.~Roshko,
	G.~Burton, M.~Colangelo, K.~Van~Tilburg, A.~Arvanitaki et~al., \emph{New
		constraints on dark photon dark matter with superconducting nanowire
		detectors in an optical haloscope},
	\href{https://doi.org/10.1103/PhysRevLett.128.231802}{\emph{Phys. Rev. Lett.}
		{\bfseries 128} 231802} (2022).
	
	\bibitem{manenti2021search}
	L.~Manenti, U.~Mishra, G.~Bruno, H.~Roberts, P.~Oikonomou, R.~Pasricha,
	I.~Sarnoff, J.~Weston, F.~Arneodo, A.~Di~Giovanni et~al., \emph{Search for
		dark photons using a multilayer dielectric haloscope equipped with a
		single-photon avalanche diode},
	\href{https://doi.org/10.1103/PhysRevD.105.052010}{\emph{Phys. Rev. D}
		{\bfseries 105} 052010} (2022).
	
	\bibitem{Caputo:2020quz}
	A.~Caputo, A.~J. Millar and E.~Vitagliano, \emph{{Revisiting Longitudinal
			Plasmon-Axion Conversion in External Magnetic Fields}},
	\href{https://doi.org/10.1103/PhysRevD.101.123004}{\emph{Phys. Rev. D}
		{\bfseries 101} 123004} (2020)
	[\href{https://arxiv.org/abs/2005.00078}{{\ttfamily 2005.00078}}].
	
	\bibitem{Gelmini:2020kcu}
	G.~B. Gelmini, A.~J. Millar, V.~Takhistov and E.~Vitagliano, \emph{{Probing
			Dark Photons with Plasma Haloscopes}},
	\href{https://doi.org/10.1103/PhysRevD.102.043003}{\emph{Phys. Rev. D}
		{\bfseries 102} 043003} (2020)
	[\href{https://arxiv.org/abs/2006.06836}{{\ttfamily 2006.06836}}].
	
	\bibitem{Mario}
	M.~G. Silveirinha, \emph{{Additional Boundary Condition for the Wire Medium}},
	\href{https://doi.org/10.1109/TAP.2006.875920}{\emph{IEEE Transactions
			Antennas and Propagation} {\bfseries 54} 1766} (2006).
	
	\bibitem{PhysRevB.101.075127}
	M.~A. Gorlach and M.~Lapine, \emph{{Boundary Conditions for the
			Effective-Medium Description of Subwavelength Multilayered Structures}},
	\href{https://doi.org/10.1103/PhysRevB.101.075127}{\emph{Phys. Rev. B}
		{\bfseries 101} 075127} (2020).
	
	\bibitem{Sajjad}
	M.~S. Mirmoosa, F.~R\"uting, I.~S. Nefedov and C.~R. Simovski,
	\emph{{Effective-Medium Model of Wire Metamaterials in the Problems of
			Radiative Heat Transfer}}, {\emph{Journal of Applied Physics} {\bfseries 115}
		234905} (2014).
	
	\bibitem{Simovski2018}
	C.~Simovski, \emph{{{Composite Media with Weak Spatial Dispersion}}}. Jenny
	Stanford Publishing, 2018.
	
	\bibitem{cst}
	\emph{{CST Studio Suite® (version 2020.03)}},  {Dassault Systèmes}.
	
	\bibitem{Belov2002}
	P.~A. Belov, S.~A. Tretyakov and A.~J. Viitanen, \emph{{Dispersion and
			Reflection Properties of Artificial Media Formed by Regular Lattices of
			Ideally Conducting Wires}},
	\href{https://doi.org/10.1163/156939302X00688}{\emph{J. Electromagn. Waves
			Appl.} {\bfseries 16} 1153} (2002).
	
	\bibitem{Maslovski2009}
	S.~I. Maslovski and M.~G. Silveirinha, \emph{{Nonlocal permittivity from a
			quasistatic model for a class of wire media}},
	\href{https://doi.org/10.1103/PhysRevB.80.245101}{\emph{Phys. Rev. B -
			Condens. Matter Mater. Phys.} {\bfseries 80} 1} (2009)
	[\href{https://arxiv.org/abs/0908.1104}{{\ttfamily 0908.1104}}].
	
	\bibitem{Olyslager2005}
	F.~Olyslager and D.~de~Zutter, \emph{{Skin Effect}}, ch.~18, pp.~4669--4675.
	\newblock John Wiley \& Sons, 2005.
	
	\bibitem{landau1995electrodynamics}
	L.~Landau, E.~Lifshitz and L.~Pitaevskii, \emph{{Electrodynamics of Continuous
			Media: Volume 8}}, Course of theoretical physics. Elsevier Science, 1995.
	
	\bibitem{pozar2011microwave}
	D.~Pozar, \emph{Microwave Engineering, 4th Edition}. Wiley, 2011.
	
	\bibitem{Ahn:2017smt}
	S.~Ahn, S.~Youn, J.~Yoo, D.~Kim, J.~Jeong, M.~Ahn, J.~Kim, D.~Lee, J.~Lee,
	T.~Seong and Y.~Semertzidis, \emph{{Magnetoresistance in Copper at High
			Frequency and High Magnetic Fields}},
	\href{https://doi.org/10.1088/1748-0221/12/10/P10023}{\emph{JINST} {\bfseries
			12} P10023} (2017) [\href{https://arxiv.org/abs/1705.04754}{{\ttfamily
			1705.04754}}].
	
	\bibitem{Darko2005}
	D.~Kajfez, \emph{{Q-Factor}}, ch.~16, pp.~3935--3947.
	\newblock John Wiley \& Sons, 2005.
	
\end{thebibliography}

\providecommand{\href}[2]{#2}\begingroup\raggedright\endgroup	

\end{document}